\documentclass[prx,superscriptaddress,twocolumn]{revtex4}

\usepackage{graphicx}
\usepackage{dcolumn}
\usepackage{array}
\usepackage{tabularx}
\usepackage{siunitx}
\usepackage{textcomp}

\usepackage{floatrow}
\usepackage[dvipsnames]{xcolor}

\usepackage{eucal}
\usepackage[dvips]{epsfig}
\usepackage{amssymb}
\usepackage{amsfonts}
\usepackage{gensymb}

\usepackage{amsmath,amsthm}
\usepackage{cleveref}
\usepackage{siunitx}
\usepackage{chngcntr}

\newcommand{\be}{\begin{eqnarray}}
\newcommand{\ee}{\end{eqnarray}}
\newcommand{\bse}{\begin{subequations}}
\newcommand{\ese}{\end{subequations}}

\newcommand{\bnum}{\begin{enumerate}}
\newcommand{\enum}{\end{enumerate}}

\newcommand{\bit}{\begin{itemize}}
\newcommand{\eit}{\end{itemize}}

\newcommand{\bc}{\begin{cases}}
\newcommand{\ec}{\end{cases}}

\newcommand{\bpm}{\begin{pmatrix}}
\newcommand{\epm}{\end{pmatrix}}

\newcommand{\bvm}{\begin{vmatrix}}
\newcommand{\evm}{\end{vmatrix}}

\newcommand{\bs}{\boldsymbol}

\newcommand{\ga}{\alpha}

\newcommand{\eps}{\epsilon}

\newcommand{\p}{\partial}
\newcommand{\f}{\frac}

\newcommand{\tn}{\textnormal}

\begin{document}

\title{Encounter rates between bacteria and small sinking particles}

\author{Jonasz S\l{}omka}
\affiliation{Institute of Environmental Engineering,
Department of
Civil, Environmental and Geomatic Engineering,
ETH Z{\"u}rich}
\author{Uria Alcolombri}
\affiliation{Institute of Environmental Engineering,
Department of
Civil, Environmental and Geomatic Engineering,
ETH Z{\"u}rich}
\author{Eleonora Secchi}
\affiliation{Institute of Environmental Engineering,
Department of
Civil, Environmental and Geomatic Engineering,
ETH Z{\"u}rich}
\author{Roman Stocker}
\affiliation{Institute of Environmental Engineering,
Department of
Civil, Environmental and Geomatic Engineering,
ETH Z{\"u}rich}
\author{Vicente I. Fernandez}
\affiliation{Institute of Environmental Engineering,
Department of
Civil, Environmental and Geomatic Engineering,
ETH Z{\"u}rich}

\date{\today}
\begin{abstract} 
Bacteria in aquatic environments often interact with particulate matter. A key example is bacterial degradation of marine snow responsible for carbon export from the upper ocean in the biological pump. The ecological interaction between bacteria and sinking particles is regulated by their encounter rate, which is therefore important to predict accurately in models of bacteria-particle interactions. Models available to date cover the diffusive encounter regime, valid for sinking particles larger than the typical run length of a bacterium. The majority of sinking particles, however, are small, and the encounter process is then ballistic rather than diffusive. In the ballistic regime, the shear generated by the particle's motion can be important in reorienting bacteria and thus determining the encounter rate, yet the effect of shear is not captured in current encounter rate models. Here, we combine analytical and numerical calculations to quantify the encounter rate between sinking particles and non-motile or motile microorganisms in the ballistic regime, explicitly accounting for the hydrodynamic shear created by the particle and its coupling with microorganism shape. We complement results with selected experiments on non-motile diatoms. We find that the shape---shear coupling has a considerable effect on the encounter rate and encounter location through the mechanisms of hydrodynamic focusing and screening, whereby elongated microorganisms preferentially orient normally to the particle surface downstream of the particle (focusing) and tangentially to the particle surface upstream of the particle (screening). We study these mechanisms as a function of the key dimensionless parameters: the ratio of particle sinking speed to microorganism swimming speed, the ratio of particle radius to microorganism length, and the microorganism's aspect ratio. We find that non-motile elongated microorganisms are screened from sinking particles in ballistic interactions because shear aligns them tangentially to the particle surface. As a result, the encounter rate is reduced by a factor proportional to the square of the microorganism aspect ratio as compared to a spherical microorganism. For motile elongated microorganisms, hydrodynamic focusing increases the encounter rate approximately twofold compared to the case without shear when particle sinking speed is similar to microorganism swimming speed, whereas for very quickly sinking particles hydrodynamic screening can reduce the encounter rate below that of non-motile microorganisms. We apply these results to predict the encounter rates of submillimetric marine particles with bacteria under natural ocean conditions. In this size range, which covers the most abundant marine particles, the shear-induced reorientation competes with randomization of the swimming direction due to Brownian effects and run-and-tumble motility. We present comprehensive maps that connect the ballistic and diffusive limits and yield the encounter rate as a function of shape, motility and particle characteristics. Overall, our results indicate that hydrodynamic shear is important to consider in the colonization and ultimately degradation of marine particles, with a direct impact on which particles are colonized and where. Shear should thus be taken into account to predict the dynamics of settling particles responsible for the large carbon flux in the ocean's biological pump.
\end{abstract}


\maketitle

\section{Introduction}
Encounters involving small particles suspended in a fluid underpin many industrial, physical and biological processes. In papermaking, too high a collision rate between cellulose fibers leads to excessive fiber flocculation and poor paper quality~\cite{Lundell2011}. In the atmosphere, precipitation formation  relies on encounters between water droplets in clouds under the combined action of gravity and turbulence~\cite{Falkovich2002}. In the ocean, encounter rates between microscopic phytoplankton following a phytoplankton bloom determine the formation of marine snow responsible for the biological pump, the vertical flux of carbon from the upper ocean to its depths~\cite{Jackson2005}. Living organisms extend the complexity of the encounter processes occurring in non-motile systems by additional mechanisms. Microorganisms and plankton dwelling in the oceans can navigate through water in search of food and motility greatly enhances the encounter rates of these microscopic organisms with resource patches~\cite{Guasto2012}. Compared to non-motile microorganisms, whose encounter rate is proportional to the low diffusivity associated with Brownian motion, motile microorganisms have a much higher (often, 100- to 1000-fold) encounter rate, since their motility effectively enhances diffusivity~\cite{Guasto2012}. Of particular importance for the biogeochemical cycles of carbon in the ocean are the encounters between bacteria and sinking particles of organic matter. Once attached to a particle, bacteria can grow on it and solubilize it, thus reducing the flux of carbon to the deep ocean~\cite{Longhurst1989,Ducklow2001}, a fundamental process in climate-relevant carbon dynamics. Accurate models of the encounter rate between bacteria and particles valid across a wide range of particle sizes are thus important to estimate the role of bacteria in the carbon pump. To date, however, encounter rate formulations have focused on the diffusive regime suitable for large particles. Here, we study the encounter between microorganisms and sinking particles in the ballistic limit, relevant for the most abundant small particles, with focus on the impact of fluid flow and the associated shear generated by the particle on the encounter rates.

Theoretical estimates of the encounter rates between microorganisms and sinking particles have thus far primarily built on modeling microorganisms as spherical colloids and motility as a diffusive process~\cite{Karp-Boss1996,Kiorboe2002}. These simplifying assumptions map the microbial encounter with particles onto the classical problem of heat and mass transfer~\cite{Friedlander_AIChEJ1957,Kiorboe2002}. By construction, this approach assumes particles are larger than the run length of a bacterium. Since the latter is of the order of tens to hundred of microns~\cite{Kiorboe2002,Son2016}, the diffusive approximation is limited to particles larger than several hundred microns. Yet, due to the power-law nature of the marine particle size spectrum, the most abundant particles in the ocean have sizes below hundred microns~\cite{Bochdansky2016}. In this increasingly ballistic regime, the coupling between the flow generated by the particle and the swimming of bacteria may dominate the bacterial orientational dynamics, in contrast to the diffusive regime~\cite{Kiorboe2001}. There is substantial experimental and theoretical evidence that fluid velocity gradients (shear) can dramatically modify the swimming trajectories of microorganisms~\cite{Marcos_PRL2009, Rusconi_NatPhys2014,Barry_JRoyalSocInt2015}. A primary mechanism is shear-induced reorientation, whereby the torque associated with fluid velocity gradients reorients microorganisms and thus impacts their swimming direction and where they end up in the flow. For example, a simple parabolic flow can lead to shear-trapping and bacterial accumulation near microchannel walls~\cite{Rusconi_NatPhys2014}. Shear-induced reorientation is a general phenomenon, applicable to any elongated bacteria that swim in flow, yet its impacts on the fundamental problem of the encounter rate between microorganisms and sinking aggregates in the ballistic range has to date not been considered. 

Here, we combine analytical and numerical calculations with experiments to study encounters between non-motile and motile microorganisms and sinking particles in the ballistic regime, with focus on how the flow created by the particle affects bacterial trajectories and ultimately the encounter rates. For the classical Stokes flow around a sphere, we show analytically that the orientational dynamics of elongated bacteria - unlike spherical particles - break the fore--aft symmetry of the flow streamlines, with major consequences on encounter rates and attachment location. Non-motile elongated bacteria orient tangentially to the particle surface as they pass by the particle, which reduces their encounter rate by a factor proportional to the square of the bacterial aspect ratio. For motile elongated bacteria, the encounter rate is very sensitive to the particle sinking speed relative to the bacterial swimming speed. When both speeds are comparable, shear increases the encounter rate about twofold and leads to preferential attachment to the leeward side of the particle. For rapidly sinking particles, shear screens motile bacteria from the sinking particle and surprisingly, the encounter rate drops far below the limit corresponding to non-motile bacteria. 

This work is organized as follows: we introduce the model of the encounter process and define the relevant observables in Section~\ref{sec:model}. To quantify the impact of shear on bacterial orientation, we classify the asymptotic configurations that ellipsoids assume in general flows and then apply the results to the Stokes flow around a sphere in Section~\ref{sec:analytical}. The encounter of non-motile and motile microorganisms with sinking particles is studied in Sections~\ref{sec:non-motile} and~\ref{sec:motile}. We discuss the biophysical consequences of our mechanistic description of the encounter process in Section~\ref{sec:discussion} and draw conclusions in Section~\ref{sec:conclusions}.

\begin{figure}[t]
  \caption{The ballistic model of the encounter between bacteria and sinking particles includes the impact of shear on bacterial trajectories. 
  (a)~Spherical particle sinks under gravity with speed given by the Stokes law~(\ref{eq:terminal_speed}) and induces the Stokes flow~(\ref{eq:Stokes_flow_sphere}) around it. Bacteria are modeled as self-propelled ellipsoids of aspect ratio $\alpha$ with the center of mass $\bs x (t)$ and the tail-to-head orientation $\bs p(t)$ obeying~Eq.~(\ref{eq:EOMbacterium}).
  (b)~Representative trajectories for a bacterium starting at $[x,y,z]=R[2,-2,-2]$ with a random initial orientation ($U/U_b=3$ and $\alpha=10$). Red (black) trajectories correspond to interceptions (misses). Interceptions are characterized by the landing position on the particle as well as the initial orientation. All such initial orientations define the interception probability starting at the given position~[red points in the inset,~Eq.~(\ref{eq:enc_prob_x})].
}
  \centering
    \includegraphics[width=1.0\textwidth]{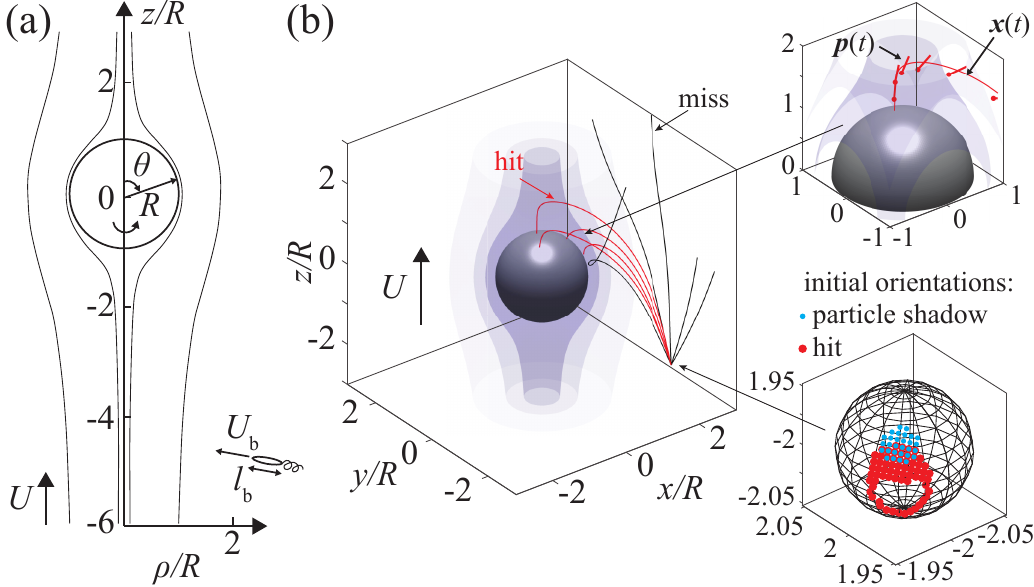}
\label{fig:sample_trajectories}
\end{figure}

\section{Model}
We model a marine snow particle as a sphere sinking in a quiescent fluid and bacteria as elongated and self-propelled ellipsoids~(Section~\ref{sec:EOM}). The encounter process is quantified through encounter rate, encounter efficiency and distribution of interception locations~(Section~\ref{sec:observables}).

\label{sec:model}
\subsection{Equations of motion}
\label{sec:EOM}
The most abundant marine snow particles in the ocean have sizes in the range up to several hundred microns~\cite{Bochdansky2016} and sinking speeds up to about a millimeter per second~\cite{Alldredge1989}, which gives Reynolds number up to about 0.1. In this viscosity-dominated regime, the gravitational and viscous forces on the particle balance, implying that a spherical particle of radius $R$ sinks at the constant terminal speed given by the Stokes law
\be
\label{eq:terminal_speed}
U=\f{2}{9}\f{\rho_{\tn{p}}/\rho_\tn{w}-1}{\nu}g R^2,
\ee
where $\rho_{\tn{p}}$ and $\rho_\tn{w}$ are the densities of the particle and water, respectively, $\nu$ is the kinematic viscosity of water and $g$ is the gravitational acceleration. In the reference frame fixed at particle and moving with it~[Fig.~\ref{fig:sample_trajectories}(a)], the flow is described by the classic Stokes flow
\be
\label{eq:Stokes_flow_sphere}
\bs v=U\cos\theta \Big(1+\f{R^3}{2 r^3}-\f{3 R}{2 r}\Big)\hat{\bs r}-U\sin\theta\Big(1-\f{R^3}{4 r^3}-\f{3 R}{4r}\Big)\hat{\bs\theta},
\ee
where $U$ is the sinking speed given by Eq.~(\ref{eq:terminal_speed}).

We model a bacterium as a small self-propelled elongated ellipsoid characterized by three parameters: length $l_\tn{b}$, aspect ratio $\alpha$ and swimming speed $U_\tn{b}$. The position and orientation of the bacterium at time $t$ are given by $\bs x(t)$ and $\bs p (t)$, where the latter (unit) vector points from the bacterial tail to its head~[Fig.~\ref{fig:sample_trajectories}(b)]. The dynamics of $\bs x$ and $\bs p$ are governed by
\bse
\label{eq:EOMbacterium}
\be
\label{eq:EOMbacteriumA}
\dot{\bs x}&=&U_\tn{b}\bs p+\bs v, \\
\label{eq:EOMbacteriumB}
\dot{\bs p}&=&(\bs I-\bs p \bs p^\tn{T})(\gamma E+W)\bs p.
\ee
\ese
Eq.~(\ref{eq:EOMbacteriumA}) states that the total bacterial velocity $\dot{\bs x}$ is a superposition of self-propulsion with speed $U_\tn{b}$ in the direction $\bs p$ and the flow $\bs v$~(\ref{eq:Stokes_flow_sphere}) around the particle. Eq.~(\ref{eq:EOMbacteriumB}) is the classic Jeffrey equation for the orientational dynamics of ellipsoids in flow~\cite{Jeffery1922}. The tensors $E$ and $W$ are the symmetric and anti-symmetric parts of the velocity gradient $A_{ij}=\p_j v_i$. The bacterial aspect ratio enters the dynamics~(\ref{eq:EOMbacterium}) through the shape parameter $\gamma=(\alpha^2-1)/(\alpha^2+1)$; it vanishes for spheres, is positive for elongated organisms and negative for oblate ones. 

\subsection{Physical observables}
\label{sec:observables}
Let $p(\bs x,\bs p)$ be the probability of an encounter between the sinking particle and a bacterium starting at the initial position $\bs x$ with head pointing in the direction $\bs p$. For the ballistic model~(\ref{eq:EOMbacterium}), $p$ is either zero or one since the initial condition $(\bs x,\bs p)$ determines a unique bacterial trajectory; when Eq.~(\ref{eq:EOMbacterium}) is supplemented with rotational diffusion, $p$ can take a range of values between 0 and 1. Averaging over random orientations yields the encounter probability $P(\bs x)$ for an initial position $\bs x$
\be
\label{eq:enc_prob_x}
P(\bs x)=\int d\bs p p(\bs x,\bs p).
\ee
Intuitively, $P(\bs x)$ is the relative solid angle extended by initial bacterial orientations that lead to the interception~(red area in the inset of~Fig.~\ref{fig:sample_trajectories}(b)].
Let  $(z,\rho,\phi)$ be the cylindrical coordinate system with origin fixed at the sinking particle. Due to rotational symmetry around the $z$-axis, we have $P(\bs x)=P(z,\rho)$. To define the encounter rate and interception efficiency, suppose that the sinking particle enters a region of uniform concentration $n$ of randomly oriented bacteria. Let $\hat{N}(t,z)$ be the total number of encounters with bacteria that at time $t$ are located at a $z$-plane below the sinking particle (upstream of the particle, $z<0$) and collide with the particle at some later time. In a short interval $(t,t+dt)$, the change in $\hat{N}$ due to the encounters with bacteria with initial positions in the thin sheet $(z,z+Udt)$ is $2\pi n  Udt \int_0^\infty P(z,\rho)\rho d\rho$. Therefore, for a constant sinking speed, the encounter rate $d\hat{N}/dt(z)$ is independent of time and is given by
\be
\label{eq:Nhatdot}
d\hat{N}/dt(z)=2\pi n  U \int_0^\infty P(z,\rho)\rho d\rho.
\ee
For a $z$-plane far away from the sinking particle $|z|\gg R$ the fluid is practically undisturbed, making it meaningful to define the $z$-independent encounter rate $\dot N=dN/dt$
\be
\label{eq:Ndot}
\dot N=d\hat{N}/dt(|z|\to \infty).
\ee
In simulations, we fix the starting plane at $z=-6R$, which amounts to making the approximation $\dot N \approx d\hat{N}/dt(z=-6R)$. To scale out the concentration $n$, we often focus on $\dot N/n$, the \lq encounter rate kernel\rq~\cite{Kiorboe2002}.

To further scale out factors intrinsic to the sinking particle, the radius $R$ and velocity $U$, we follow the notation used in filtration literature and define the dimensionless interception efficiency $\eta$ as the ratio of volume cleared and volume swept by the particle~\cite{Friedlander_AIChEJ1957}
\be
\label{eq:eta}
\eta=\f{\dot N/n}{\pi R^2 U}=\f{2}{R^2}\int_0^\infty P(\rho)\rho d\rho.
\ee
Intuitively, $\eta=1$ means that the sinking particle collects bacteria from a volume of water equal to the volume of the cylinder the particle sweeps. For small non-motile colloids, we expect $\eta \ll 1$ because the colloids are constrained to the flow streamlines, which limits the interception to a narrow region near the particle centerline, the \lq stagnation line\rq~[Fig.~\ref{fig:nonmotile}(a) and Section~\ref{sec:non-motile_bac}].

In addition to computing the encounter rate and encounter efficiency, we will quantify the location on the sinking particle where the bacteria land. Let $\xi(\theta,\phi)$ be the distribution of the interception locations, where $\theta$ and $\phi$ are the colatitude and the azimuth coordinates on the particle, respectively. We normalize $\xi(\theta,\phi)$ as the probability density function over the unit sphere, $
\int_0^\pi\int_0^{2\pi} \xi(\theta,\phi) d\Omega=1$, where $d\Omega=\sin\theta d\theta d\phi$. Rotational symmetry implies that
$\xi(\theta,\phi) = \xi(\theta).$ Finally, the mean interception colatitude is
\be
\langle\theta\rangle=\int_0^\pi\int_0^{2\pi} \theta\xi(\theta,\phi) d\Omega=
2\pi\int_0^\pi\theta\xi(\theta) \sin\theta d\theta.
\ee
For example, $\ang{0}<\langle\theta\rangle<\ang{90}$~(northern hemisphere, downstream) implies preferential leeward attachment, while $\ang{90}<\langle\theta\rangle<\ang{180}$~(southern hemisphere, upstream) indicates attachment to the front.

\begin{figure}[t!]
  \caption{
In flow, spherical bacteria rotate around the vorticity vector~(a), whereas elongated or oblate ones eventually point in the direction of the largest effective deformation rate~(b) or rotate in a certain plane~(c). (a-c)~Phase portraits of the Jeffrey Eq.~(\ref{eq:JeffreysEq2}) for the bacterial tail-head vector $\bs p(t)$ reorienting under the velocity gradient $A$ for different $A$ and aspect ratios~$\ga$.
(a)~Spherical bacteria always rotate around the vorticity $\bs\omega$~($z$-axis).
(b,c)~Depending on the flow being strain- or rotation-dominated, elongated bacteria (or flat disks) eventually point along the direction of the largest effective deformation rate~[$z$-axis in (b)] or rotate in the plane perpendicular to the real eigenvector of $(A^\gamma)^\tn{T}$~[equator in (c)]. 
(d) Time series of the components of $\bs p(t)$ for the case shown in (c). The rotation in the $x-y$ plane is nonuniform: the rod accelerates when approaching the straining direction but slows down when near the axis of compression; the rotation frequency is given by the imaginary part of the complex eigenvalue of $A^\gamma$. Parameters: $A=[0,-1/2,0;1/2,0,0;0,0,0], \ga=1$~(a), $A=[-3/4,0,0;0,-1/4,0;0,0,1], \ga=\infty$~(b) and $A=[1/2,1,0;-1/4,1/2,0;0,0,-1], \ga=\infty$~(c).
}
  \centering
    \includegraphics[width=1.0\textwidth]{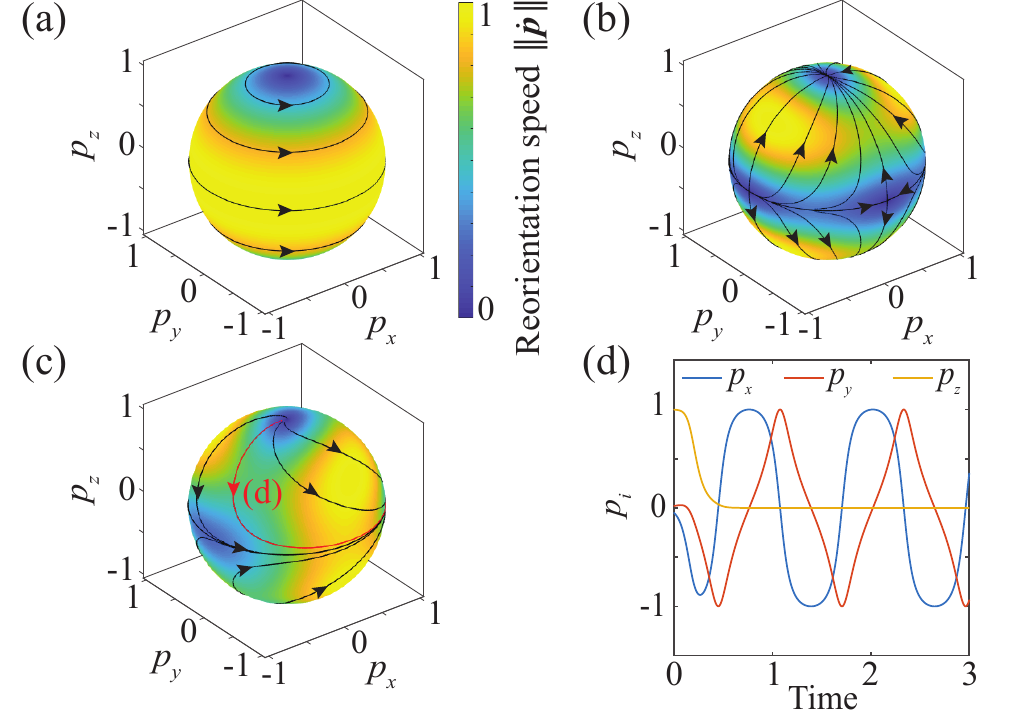}
\label{fig:Jeff_asymp}
\end{figure}

\section{Ellipsoids in flow}
\label{sec:analytical}
Shear preferentially reorients rods and disks, such as elongated bacteria or flat diatoms, along certain directions. Depending on the flow being strain- or rotation-dominated, ellipsoids eventually point in the direction of the largest deformation rate or rotate in a certain plane~(Section~\ref{sec:Jeffrey_asymp}). Applying this classification to the Stokes flow induced by the sinking particle reveals that ellipsoids - unlike spheres - break the fore-aft symmetry of the flow streamlines~(Section~\ref{sec:Stokesflow_asymp}). This quasi-static picture will be essential to rationalize the subsequent simulations of the encounter problem as it underpins the phenomena of hydrodynamic focusing and screening.

\subsection{Ellipsoids in general flows}
\label{sec:Jeffrey_asymp}
The asymptotic orientation of a nonspherical microorganism held fixed in flow but free to reorient under the action of the velocity gradient $A$ follows from the long-time limit of the Jeffrey Eq.~(\ref{eq:EOMbacteriumB}). Previous studies focused on special cases with $A$ derived from, for example, simple shear or rotational flows; in the former case the dynamics collapse onto one of the many degenerate limit cycles, the well-known Jeffrey orbits~\cite{Jeffery1922,Junk_JMFM2007}. For a random $A$, neglecting marginal cases, two scenarios are possible: either a rod asymptotically points towards the direction of the largest effective deformation rate or it rotates in a certain plane. The first possibility has been known~\cite{Junk_JMFM2007} and corresponds to the rate of strain $E$ out-competing the rate of rotation $W$. The second scenario complements the study in~\cite{Bretherton1962} and generalizes the Jeffrey orbits to generic rotational flows and arises when $W$ dominates over $E$. Detailed derivations are given in the Appendices~\ref{Appsec:Jeff_linearstability} and~\ref{Appsec:Jeff_limitcycle}. 

The velocity gradient tensor $A$ has nine components, eight of which are independent for an incompressible flow~(since $A_{ii}=\p_i v_i=0$). The symmetric part of $A$, the rate of strain $E$, describes the rate at which the fluid stretches and compresses~\cite{Batchelor1979}. The antisymmetric part $W$ represents the fluid rate of rotation and is determined by the vorticity $\bs \omega=\nabla\times\bs u$ as  $W_{ij}=-\f{1}{2}\eps_{ijk}\omega_k$. Given $A$, it is the weighted sum $A^\gamma=\gamma E+W$ that enters the Jeffrey Eq.~(\ref{eq:EOMbacteriumB}), where $\gamma$ is the shape parameter $\gamma=(\alpha^2-1)/(\alpha^2+1)$ determined by the organism aspect ratio $\alpha$. In this notation, the Jeffrey Eq.~(\ref{eq:JeffreysEq2}) reads
\be
\label{eq:JeffreysEq2}
\dot{\bs p}&=&(\bs I-\bs p \bs p^\tn{T})A^\gamma\bs p.
\ee
Eq.~(\ref{eq:JeffreysEq2}) is a dynamical system on the unit sphere of orientations~(Fig.~\ref{fig:Jeff_asymp}). We first discuss the case of spherical microorganisms ($\alpha=1,\gamma=0$) and then describe in detail the response of elongated bacteria ($\alpha>1, \gamma>0$); the case of oblate microorganisms~ ($\alpha<1, \gamma<0$) is dual to that of elongated ones.

For spherical microorganisms, the shape parameter vanishes ($\gamma=0$) and Eq.~(\ref{eq:JeffreysEq2}) simplifies to
\be
\label{eq:JeffEq_spheres}
\dot{\bs p}&=&\f{1}{2}\bs\omega\times\bs p,
\ee
where $\bs\omega$ is the vorticity. Thus, spherical microorganisms respond to the fluid rotation but are unaffected by the fluid straining motion. Eq.~(\ref{eq:JeffEq_spheres}) can be solved exactly in this case~\cite{Junk_JMFM2007}: for a given initial orientation $\bs p(0)$, the solutions of Eq.~(\ref{eq:JeffEq_spheres}) correspond to $\bs p(t)$ rotating around the vorticity vector $\bs \omega$ and in the same sense as $\bs \omega$~[Fig.~\ref{fig:Jeff_asymp}(a)]. The angle between $\bs p(t)$ and $\bs\omega$ is fixed by the initial orientation and the rotation rate is $\|\bs\omega\|/2$. Equivalently, the solutions to Eq.~(\ref{eq:JeffEq_spheres}) on the unit sphere of orientations with $\bs \omega$ pointing along the $z$-axis are given by the circles of a fixed latitude.

Elongated or oblate microorganisms respond to both, the fluid rate of strain and rate of rotation. In this case, it appears impossible to find analytical solutions to Eq.~(\ref{eq:JeffreysEq2}); instead, standard dynamical system theory helps to identify the long-time response. Since the fixed points of~Eq.~(\ref{eq:JeffreysEq2}) are given by the real eigenvectors of $A^\gamma$~\cite{Junk_JMFM2007}, it is the eigendecomposition of $A^\gamma$ that determines the asymptotic response. For a random $A^\gamma$, two cases are possible: either $A^\gamma$ has three real eigenvalues or one real eigenvalue and two complex conjugate eigenvalues. In the first case, the eigenvector corresponding to the largest positive eigenvalue is an attractive fixed point of~(\ref{eq:JeffreysEq2})~[Fig.~\ref{fig:Jeff_asymp}(b)]. In the second case, when the real eigenvalue is positive, the corresponding eigendirection is still attractive, but once this eigenvalue is negative, the eigendirection becomes unstable and the dynamics collapse onto a limit cycle [Fig.~\ref{fig:Jeff_asymp}(c)]. We next discuss these asymptotic scenarios in more detail.

Let $\lambda_i$ and $\bs \lambda_i$, where $i=1,2,3$, be the eigenvalues and eigenvectors of $A^\gamma$. The incompressibility of the flow requires that $\lambda_1+\lambda_2+\lambda_3=0$. When all $\lambda_i$'s are real, the Jeffrey Eq.~(\ref{eq:JeffreysEq2}) has three pairs of fixed points corresponding to $\bs p=\pm \bs \lambda_i$. Assuming that $\lambda_1<\lambda_2<\lambda_3$, the fixed points are respectively: a repulsive node ($\bs \lambda_1$), a saddle ($\bs \lambda_2$), and an attractive node ($\bs \lambda_3$). As a consequence, a random initial orientation eventually collapses onto the stable direction $\bs p=\pm\bs \lambda_3$~ [Fig.~\ref{fig:Jeff_asymp}(b)]. The timescale $\tau_{\bs\lambda_3}$ associated with this reorientation is estimated as inverse of the average of the eigenvalues of the linearized version of Eq.~(\ref{eq:JeffreysEq2}) near the fixed point $\bs\lambda_3$ and is given by $\tau_{\bs\lambda_3}^{-1}=\f{3}{2}\lambda_3$. Since the system orients along $\pm\bs \lambda_3$, the case when all $\lambda_i$'s are real corresponds to the rate of strain $E$ outcompeting the rate of rotation $W$.

When $A^\gamma$ has a pair of complex eigenvalues $\lambda_1$ and $\lambda_1^*$, and a real eigenvalue $\lambda_3$, the eigenvectors corresponding to $\{\lambda_{1}, \lambda_1^*\}$ are complex, implying that there are only two fixed points given by $\bs p=\pm \bs\lambda_3$. When $\lambda_3>0$~(stretching), the fixed point $\bs \lambda_3$ is an attractive spiral and represents the asymptotic direction. The appearance of complex eigenvalues signals the raising importance of the rate of rotation $W$, but when $\lambda_3>0$, straining still dominates the response. However, when $\lambda_3<0$~(compression), the fixed point $\bs \lambda_3$ becomes a repulsive spiral and a stable limit cycle emerges. The timescale $\tau_{\bs\lambda_3}$ of spiraling onto or away from $\bs\lambda_3$ is $\tau_{\bs\lambda_3}^{-1}=\f{3}{2}|\lambda_3|$. The limit cycle corresponds to a great circle; the circle lies in the plane with normal direction given by $\bs \lambda_3'$, the eigenvector of the transpose  matrix ${A^\gamma}^{\tn{T}}$ with eigenvalue $\lambda_3$. Thus, when $\lambda_3<0$, the asymptotic state of~Eq.(\ref{eq:JeffreysEq2}) corresponds to $\bs p$ rotating in the plane normal to $\bs \lambda_3'$~[Fig.~\ref{fig:Jeff_asymp}(c)]. The angular frequency of the rotation is given by the imaginary part of the complex eigenvalue $\lambda_1$~[Fig.~\ref{fig:Jeff_asymp}(d)].

The above analysis applies to a velocity gradient $A^\gamma$ under the assumption that all its eigenvalues are different; a separate analysis is required in the degenerate case. We next study $A$ derived from the Stokes flow around a sinking sphere. 

\begin{figure*}[t!]
  \caption{Nonspherical microorganisms break the fore-aft symmetry of the Stokes flow around a sinking particle as revealed by their asymptotic orientation when held fixed in the flow. The velocity gradient $A$~[Eq.~(\ref{eq:VGspherical_ortho})] determines the local long-time orientation of the bacterial tail-to-head vector~$\bs p(t)$~(Fig.~\ref{fig:Jeff_asymp}). 
  (a)~Spherical bacteria respond symmetrically upstream and downstream of the particle: $\bs p$ rotates around the vorticity~$\bs \omega\propto\hat{\bs\phi}$ and in the same sense as $\bs\omega$~(arrow); the broken line indicates that there are many possible orbits~[Fig.~\ref{fig:Jeff_asymp}(a)]. 
(b)~Perfect rods exhibit three regions of different asymptotic orientations. In regions I and III the rate of strain outcompetes the vorticity~[Fig.~\ref{fig:Jeff_asymp}(b)], whereas the vorticity dominates in the region II~[Fig.~\ref{fig:Jeff_asymp}(c)]. Specifically, in the upstream region I, bacteria eventually point along the azimuth $\bs p=\pm\hat{\bs\phi}$. The two subregions inside the region I differ only by how the asymptotic orientation is attained~(attractive node vs spiral). In region II, $\bs p$ eventually rotates in the plane perpendicular to $\hat{\bs\phi}$~[Fig.~\ref{fig:Jeff_asymp}(c)]. The color code shows the rotation rate normalized by $\|\bs\omega\|/2$; the sense of rotation is the same as $\bs\omega$~(solid line and arrow). In the downstream region III, $\bs p$ orients along the director field~(white lines) defined by $\bs \lambda_1$ in Eq.~(\ref{eq:Aeigvecs}).
(c)~For perfect disks, the response is a reflection of the case of perfect rods.
(d)~Ratio between the advective and reorientation timescales, $\tau_\tn{a}$ and $\tau_\tn{r}$. For $\tau_\tn{a}/\tau_\tn{r}\gg 1$, we expect non-motile microorganisms advected by the flow to follow the quasi-static reorientation effects described in~(b,c); these reorientation effects are strongest near the sinking particle.
}
  \centering
    \includegraphics[width=1.0\textwidth]{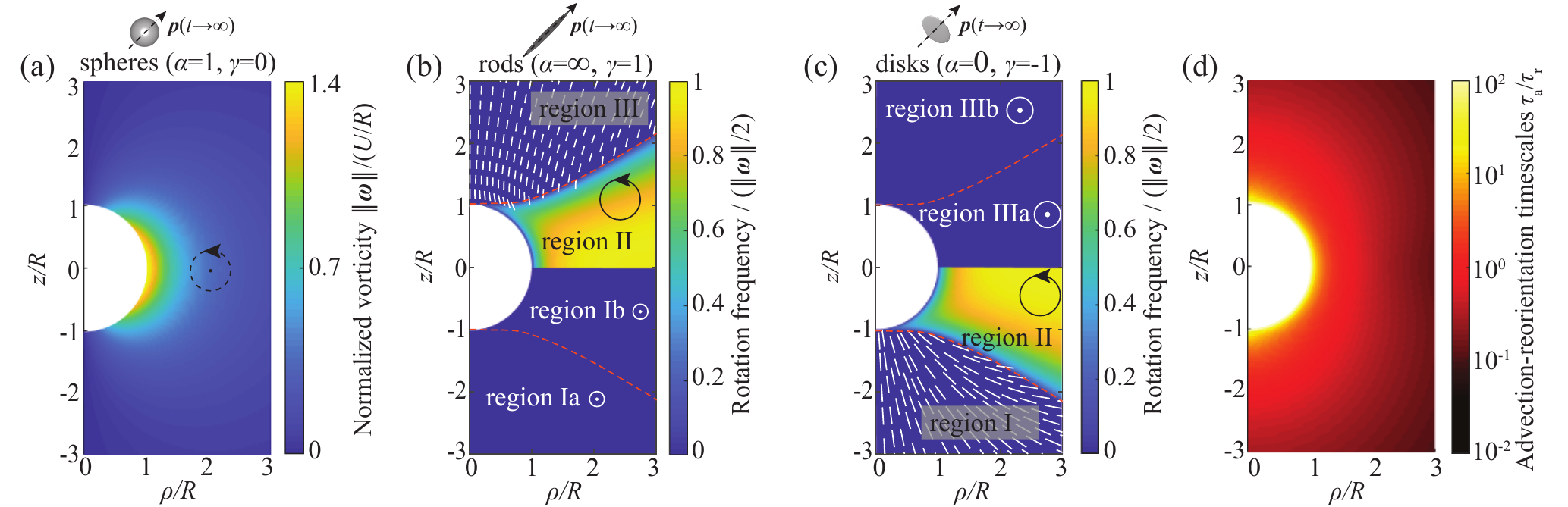}
\label{fig:particles_fixed_StokesFlow}
\end{figure*}

\subsection{Ellipsoids in the Stokes flow}
\label{sec:Stokesflow_asymp}
The above classification of the orientational response of a microorganism is now specified to the velocity gradient derived from the Stokes flow~(\ref{eq:Stokes_flow_sphere}). Physically, we describe a bacterium rotating freely under shear but with the center of mass fixed at some position. Self-propulsion and advection are still not included and this simplification makes analytical progress possible. The long-time orientation depends on bacterial position: spherical bacteria respond identically upstream and downstream of the sinking particle~[Fig.~\ref{fig:particles_fixed_StokesFlow}(a)], whereas nonspherical ones break the streamline fore--aft symmetry~[Fig.~\ref{fig:particles_fixed_StokesFlow}(b,c)]. This symmetry breaking leads to hydrodynamic focusing and screening, crucial shear-induced mechanisms that impact the full encounter problem.

Spherical bacteria respond solely to the fluid vorticity (Section~\ref{sec:Jeffrey_asymp}), which for the Stokes flow~(\ref{eq:Stokes_flow_sphere}) reads
\be
\label{eq:Stokes_vorticity}
\bs \omega=-\f{3}{2}UR \f{\sin\theta}{r^2}\hat{\bs\phi}.
\ee
It follows from Eq.~(\ref{eq:Stokes_vorticity}) that bacteria rotate around the azimuth $\hat{\bs\phi}$,~[Fig.~\ref{fig:particles_fixed_StokesFlow}(a)]. The rotation rate decays with the square of the distance from the particle; it is strongest to the side of the particle, near the equator $\theta=\pi/2$ and vanishes near the stagnation lines $\theta=0$ and $\theta=\pi$. In particular, the response of spherical bacteria preserves the fore-aft symmetry of the flow streamlines: at a fixed distance $r$, the bacterial rotation is identical at colatitudes $\theta$ and $\pi-\theta$. As discussed next, this symmetry is broken for elongated or oblate microorganisms.

We focus on perfect rods of infinite aspect ratio $\alpha\to\infty$, for which the shape factor $\gamma=1$ in the Jeffrey Eq.~(\ref{eq:EOMbacteriumB})~[Fig.~\ref{fig:particles_fixed_StokesFlow}(b)]. Note that moderate elongation gives $\gamma$ close to unity: for an aspect ratio $\alpha=10$, $\gamma\approx 0.98$. The analysis of the response of disks is dual to that of rods and we only state the results~[Fig.~\ref{fig:particles_fixed_StokesFlow}(c)]. For perfect rods, the rates of strain and rotation, $E$ and $W$, are weighted equally in the Jeffrey Eq.~(\ref{eq:JeffreysEq2}), which reduces to ($A^{\gamma=1}=A$)
\be
\dot{\bs p}&=&(\bs I-\bs p \bs p^\tn{T})A\bs p,
\ee 
where $A$ is the velocity gradient derived from the Stokes flow~(\ref{eq:Stokes_flow_sphere}). For brevity, we take $U=1$ and $R=1$. In spherical coordinates $\{r,\theta,\phi\}$ and in the usual basis of unit vectors $\{\hat{\bs r},\hat{\bs \theta},\hat{\bs\phi}\}$, the entries of $A$ read (Appendix~\ref{Appsec:AStokesFlow}, see also~\cite{Kiorboe1999})
\be
\label{eq:VGspherical_ortho}
A_{ij}=
F(r,\theta)
\begin{bmatrix}
       2 & \tan\theta & 0           \\
       -\beta(r)\tan\theta &  -1           & 0 \\
       0           & 0 &  -1
\end{bmatrix},
\ee
where $
F=3\cos\theta(r^{-2}-r^{-4})/4$ and $\beta=(r^2+1)/(r^2-1)$.
To classify the response of rods in the manner outlined in Section~\ref{sec:Jeffrey_asymp}, we find the eigenvalues of~$A$
\be
\label{eq:Aeigs}
\lambda_{1,2}=\f{F}{2}\Big(1\pm \sqrt{9-4 \beta \tan^2\theta}\Big), \quad \lambda_3=-F,
\ee
and the corresponding eigenvectors
\be
\label{eq:Aeigvecs}
\bs \lambda_{1,2}&=&[1,\f{-3\pm\sqrt{9+4 \beta\tan^2\theta}}{2\tan\theta},0],  \quad
\bs \lambda_3=\hat{\bs\phi}.
\ee
From the sing change under the square root in~Eq.~(\ref{eq:Aeigs}), it follows that the regions in the fluid in which $A$ has three real eigenvalues or a pair of complex eigenvalues plus a real eigenvalue are separated by two surfaces of revolution defined by~[broken red lines in Figs.~\ref{fig:particles_fixed_StokesFlow}(b,c)]
\be
\label{eq:complex_eigs_surface}
r^2(\theta)=(9+4\tan^2\theta)/(9-4\tan^2\theta).
\ee
Furthermore, in the region with complex eigenvalues, the real eigenvalue, $\lambda_3=-F(r,\theta)$, changes sign from negative to positive at the plane $\theta=\pi/2$, which contains the sinking particle's equator. Physically, the sign change reflects the transition of $\bs\lambda_3=\hat{\bs\phi}$ from being the direction of fluid expansion to compression as the fluid parcels travel from the southern to the northern hemisphere. The surfaces~(\ref{eq:complex_eigs_surface}) and $\theta=\pi/2$ divide the space outside the particle into three regions I, II and III~[Fig.~\ref{fig:particles_fixed_StokesFlow}(b)]. Region I is the bottom-half of the entire domain, below the equator plane $\theta=\pi/2$ and upstream of the sinking particle. It is composed of two subregions, Ia and Ib, separated by the surface~(\ref{eq:complex_eigs_surface}) [lower broken red line in Fig.~\ref{fig:particles_fixed_StokesFlow}(b)]. In Ia, all the eigenvalues are real, in Ib, there is a pair of complex eigenvalues and a positive real eigenvalue $\lambda_3>0$. In both subregions, the rate of strain dominates over the rate of rotation and the asymptotic stable direction is given by the eigenvector $\bs\lambda_3=\hat{\bs\phi}$, which always points along the azimuth. The two subregions differ only in the manner this asymptotic orientation is approached: in Ia, the convergence is overdamped~[as in Fig.~\ref{fig:Jeff_asymp}(b)] since $\hat{\bs\phi}$ is an attractive node, in Ib, the convergence is underdamped with the bacterium spiraling down onto $\hat{\bs\phi}$ since this fixed point is an attractive spiral~[as in Fig.~\ref{fig:Jeff_asymp}(c) but with arrows reversed]. This change in the nature of the convergence of $\bs p$ onto $\hat{\bs\phi}$ indicates the increasing role of vorticity near the particle equator, but $\hat{\bs\phi}$ remains the attractive fixed point in region I because the fluid has to expand along the azimuth to accommodate the sinking sphere in that region. The timescale $\tau_{\tn{I}}$ associated with convergence onto $\hat{\bs\phi}$ in region I is $\tau_{\tn{I}}^{-1}=\f{3}{2}\lambda_3=-\f{3}{2}F.$

Region II lies to the side of the particle, in between the equator plane and the surface~(\ref{eq:complex_eigs_surface}) and is the only region in which the rotation rate out-competes the rate of strain. In this region, $\lambda_{1,2}$ are complex and $\lambda_3<0$; physically, the fluid is being compressed along the azimuth as it is rolling over the particle surface due to the no-slip boundary conditions. The analysis in Section~\ref{sec:Jeffrey_asymp} implies that the rods eventually rotate in the plane orthogonal to the azimuth with frequency $F/2 \sqrt{4\beta\tan^2\theta-9}$. Thus, rods orient orthogonal to the vorticity $\bs \omega$ and rotate in the same sense as $\bs \omega$, but the rotation period is longer from the rotation rate of the fluid~[color code in region II in Fig.~\ref{fig:particles_fixed_StokesFlow}(b)]. The timescale  $\tau_{\tn{II}}$ associated with the reorientation from pointing along $\hat{\bs\phi}$ to rotating in the plane perpendicular to $\bs \phi$ is given by $\tau_{\tn{II}}^{-1}=\f{3}{2}|\lambda_3|=\f{3}{2}F$.

Region III lies downstream of the particle, above the surface~(\ref{eq:complex_eigs_surface}). Here, the strain once again dominates over rotation, but this time the asymptotic direction of rods in flow is given by the eigenvector $\bs \lambda_1$~[the white director field lines in~Fig.~\ref{fig:particles_fixed_StokesFlow}(b)]. Importantly, just behind the particle, for small colatitudes $\theta$, Eq.~(\ref{eq:Aeigvecs}) clearly predicts that the stable orientation is approximately the radial direction $\bs\lambda_1\approx [1,0,0]$. The timescale  $\tau_{\tn{III}}$ associated with the reorientation from rotating in the plane perpendicular to $\bs \phi$ in region II to pointing along $\bs \lambda_1$ is $\tau_{\tn{III}}^{-1}=\f{3}{2}\lambda_1=\f{3}{4}F\Big(1+ \sqrt{9-4 \beta \tan^2\theta}\Big)$.

The response of perfect disks ($\ga=0$, $\gamma=-1$) is dual to the case of perfect rods since $A^{\gamma=-1}=-E+W=-A^\tn{T}$. For brevity, we only summarize the results, which are essentially an upside-down version of the responds of rods~[Fig.~\ref{fig:particles_fixed_StokesFlow}(c)]. Upstream of the sinking particle, disks tend to be oriented almost tangentially to the particle, with their symmetry axis pointing in the nearly radial direction [region I, white director lines in~Fig.~\ref{fig:particles_fixed_StokesFlow}(c)]. To the side of the particle, disks rotate, with their axis of symmetry spinning in the $r-\theta$-planes (region II). Downstream of the particle (region III), disks preferentially align with the $r-\theta$ planes with their axis of symmetry pointing along the azimuth. As for rods, the region III is divided into two subregions: in IIIa, $\hat{\bs\phi}$ is an attractive spiral, in IIIb it is an attractive node.

The splitting of the fluid flow into the regions shown in~Figs.~\ref{fig:particles_fixed_StokesFlow}(b,c) is a quasi-static characterization of the dynamical system~(\ref{eq:EOMbacterium}), with microorganisms held fixed at a given position in the flow. However, a non-motile microorganism that is advected by the flow may be significantly displaced during the time it takes to achieve a given asymptotic orientation. To get further insight into Eq.~(\ref{eq:EOMbacterium}), we compare the two timescales characterizing the advection and shear-induced reorientation. For simplicity, we combine the three timescales $\tau_\tn{I,II,III}$ associated with convergence onto the asymptotic solutions to~Eq.~(\ref{eq:EOMbacteriumB}) into a single reorientation timescale $\tau_\tn{r}^{-1}\sim F\sim\|A\|_2$. We estimate the advective timescale~$\tau_\tn{a}\sim R/\|\bs v\|$ at a given position as the time needed to travel the distance $R$ at the local speed~$\|\bs v\|$.
Fig.~\ref{fig:particles_fixed_StokesFlow}(d) shows the ratio $\tau_\tn{a}/\tau_\tn{r}$ as a function of the position. In particular, in the bright oval near the particle $\tau_\tn{a}>\tau_\tn{r}$, indicating that, in that region, non-motile bacteria advected by the flow have enough time to orient under the fluid forces in the manner outlined in~Figs.~\ref{fig:particles_fixed_StokesFlow}(b,c) for immobilized bacteria.

Finally, we note that the eigenvalues of the velocity gradient $A$~(\ref{eq:VGspherical_ortho}) on the stagnation line ($\theta=0,\pi$) and the particle surface ($r=1$) have multiplicity greater than one. In this case, the analysis of Section~\ref{sec:Jeffrey_asymp} does not directly apply, yet these special locations are important for the encounter process of non-motile microorganisms, which can only approach the sinking particle near the stagnation line $\theta=\pi$. In Appendix~\ref{sec:Stokesflow_asymp_critical}, we show that on the upstream stagnation line ($\theta=\pi$) rods align tangentially to the sinking particle, while on the downstream stagnation line ($\theta=0$), rods align vertically. This picture can be inferred from Fig.~\ref{fig:particles_fixed_StokesFlow}(b) by taking the limit $\rho\to 0$. Similarly, disks align tangentially to the particle surface for $\theta=\pi$ (with axis of symmetry in the vertical direction), while they lie in the $r-\theta$ plane for $\theta=0$. Therefore, non-motile rods or disks approaching the sinking particle along the $\theta=\pi$ stagnation line orient with their longer dimension tangential to the particle surface. Furthermore, on the particle surface, shear maintains to zeroth order the tangential orientation of rods and disks as they are advected around the sinking particle~(Appendix~\ref{sec:Stokesflow_asymp_critical}). This suggests that it is the shorter dimension of rods and disks that determines their collision with the particle. However, for any finite size microorganism, one must step away from the stagnation line and the particle surface. In the vicinity of these degenerate sets, the response is captured in~Figs.~\ref{fig:particles_fixed_StokesFlow}(b,c). The asymptotic orientations rods and disks assume in their respective regions I suggest that the tangential orientation prevails. However, in regions II and III shear reorients rods and disks away from the tangential orientation. Given the size of the regions II and III for rods and disks, this reorientation should be stronger for disks, since disks experience it over a larger part of the particle surface. In the next section, we use numerical simulations to confirm this intuition: the collision radius of rods is determined by their width, not length, whereas for disks the collision radius is determined by the longest dimension.


\section{non-motile bacteria and diatoms}
\label{sec:non-motile}

Understanding the interception of non-motile elongated and oblate microorganisms by a sinking particle is important for two reasons. First, in its own right, because many marine microorganisms including many bacteria and phytoplankton species are non-motile and come in a variety of shapes, with bacteria often being spherical or elongated and phytoplankton being either elongated (e.g., chains), spherical or disk-lake (e.g., diatoms). Second, the non-motile case corresponds to the high sinking speed limit $U/U_b\to \infty$ for motile bacteria. We predict drastically different encounter rates for rods and disks: rods are particularly inefficient at intercepting the sinking particle due shear, which aligns them in the direction tangential to the particle surface~(Section~\ref{sec:non-motile_bac}). Conversely, disks eventually tumble under shear and explore their long axis to reach the collector. Experiments on elongated diatoms support this picture~(Section~\ref{sec:experiments}).

\begin{figure}[t!]
 \centering
\includegraphics[width=1.0\textwidth]{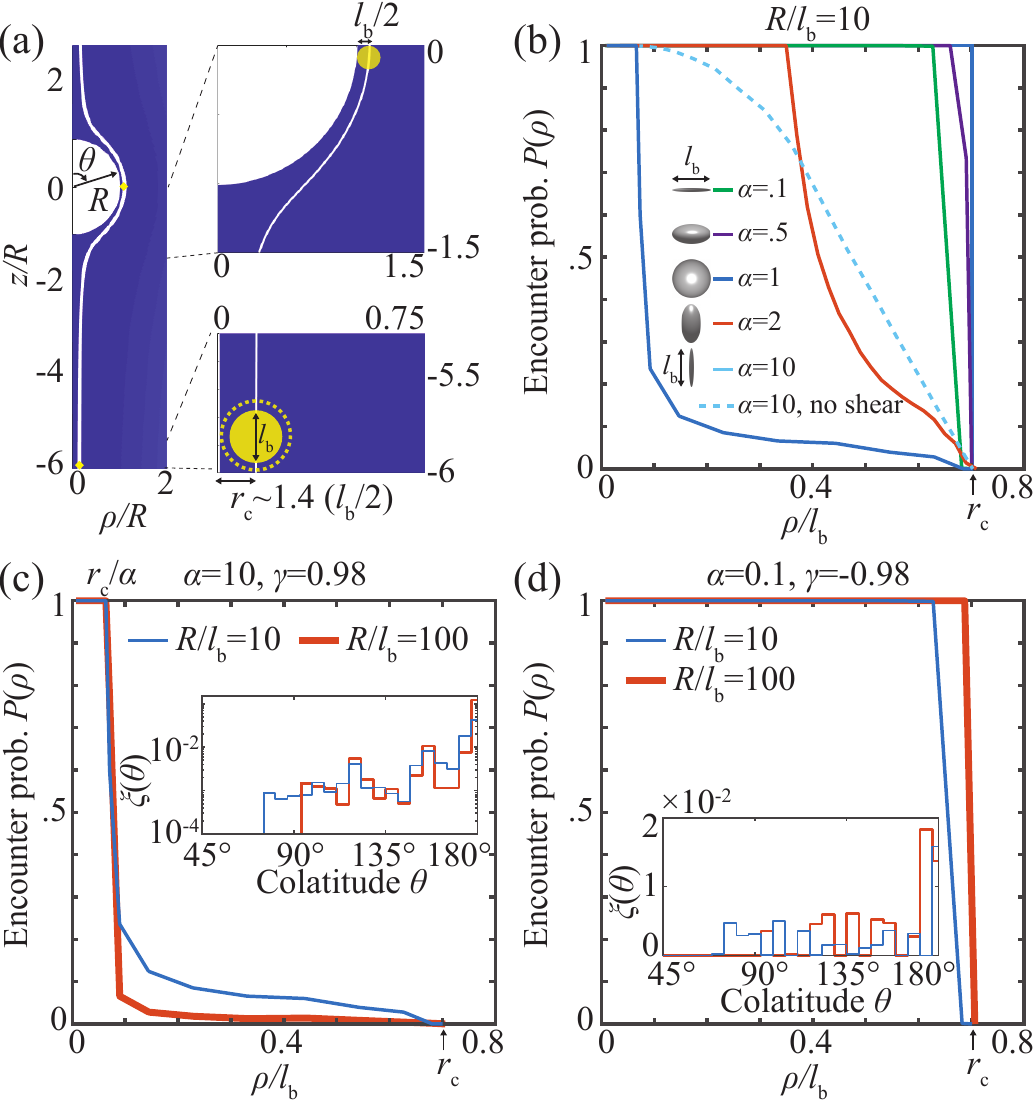}
\caption{
Generalizing the classical ballistic interception problem to nonspherical colloids reveals that shear orients non-motile elongated microorganisms tangentially to the sinking particle surface, while oblate organisms tumble near the particle. (a)~For spherical organisms, the effective collision radius $r_\tn{c}$ is defined by the critical streamline beyond which no interception occurs (white line). (b)~Encounter probability $P(\rho)$ as a function of distance $\rho$ from the origin in the initial plane $z=-6R$ for different aspect ratios~$\alpha$. For spheres and disks, $P(\rho)$ drops from 1 to 0 near $r_\tn{c}$, whereas for rods the drop occurs near $r_\tn{c}/\ga$. (c) The probability $P(\rho)$ for rods ($\ga=10$) for two different sizes $R/l_\tn{b}=10,100$ shows that the probability tail between $r_\tn{c}/\ga$ and $r_\tn{c}$ vanishes as the sinking particle size grows (or the rod becomes smaller). The tail arises to due rare tumbling events to the leeward side of the particle~(Movie~2), as can be seen from the distribution of the interception colatitudes $\xi(\theta)$ (inset). (d) Conversely, $P(\rho)$ for $\ga=0.1$ for $R/l_\tn{b}=10,100$ demonstrates that disks tumble very often - as the particle grows, $P(\rho)$ approaches a step function at $r_\tn{c}$.
}
\label{fig:nonmotile}
\end{figure}

\subsection{Interception of non-motile bacteria}
\label{sec:non-motile_bac}
The ballistic interception of non-motile microorganisms by a sinking particle is conceptually identical to the classical problem of filtration, in which a colloid is captured by a large collector~\cite{Friedlander_AIChEJ1957}. Previous works focused on spherical colloids; through numerical simulations, we extend these results to nonspherical colloids. Rods and disks, such as certain species of bacteria or diatoms, have drastically different effective collision radii. Due to shear-induced reorientation, the collision radius for a rod is determined by its width rather than length, while disks explore their full size to intercept the collector.

First, we briefly review the classical interception of small non-motile spherical beads by a large spherical collector. The reorientation of beads under the flow does not affect the interception problem, which reduces to identifying the streamline of closest approach~[Fig.~\ref{fig:nonmotile}(a)]. For the Stokes flow~(\ref{eq:Stokes_flow_sphere}), the stream-function is given by
\be
\label{eq:StokesSF}
\psi(r,\theta)=\f{1}{2}U r^2\Big(1-\f{3}{2}R/r+\f{1}{2}(R/r)^3\Big)\sin^2\theta.
\ee
The streamlines determined by Eq.~(\ref{eq:StokesSF}) have the fore--aft symmetry, which implies that the critical streamline separating captures from misses is defined by the point $r=R+l_\tn{b}/2$ and $\theta=\pi/2$. Tracing the streamline upstream from this point to $z=-6R$, where we start the simulations, defines the effective collision radius $r_\tn{c}$. For a spherical bacterium with $\l_b/R=1/10$, the collision radius is $r_\tn{c}\approx 1.4 (l_\tn{b}/2)$; the additional $40\%$ arise due to the squeezing of the streamlines near the collector. Had we traced the streamline all the way to $z\to -\infty$, the prefactor would change from 1.4 to $1.2$~\cite{Friedlander_AIChEJ1957}. In general, $r_\tn{c}$ depends very weakly on the size of spherical bacteria $l_\tn{b}/R$ and the formula $r_\tn{c}\approx 1.4 (l_\tn{b}/2)$ works very well for the bacterial sizes in the range $0<\l_b/R<1/10$. Finally, for spherical colloids, the effective collision radius and the encounter efficiency~[Eq.~(\ref{eq:eta})] are related as
\be
\label{eq:eta_spheres}
\eta_\tn{spheres}=(r_\tn{c}/R)^2.
\ee
We now turn to nonspherical organisms, for which the orientational dynamics can no longer be neglected.

As nonspherical microorganisms follow the streamlines, they can intercept the sinking particle using either their shorter or longer dimension. Two extreme scenarios are possible: an organism always aligns its longer side tangentially or perpendicular to the particle, which modifies its collision efficiency by a factor of~$\ga^2$. Assuming negligible rotational diffusion, shear impacts the orientation of nonspherical organisms only through the aspect ratio $\ga$. This follows from nondimensionalizing Eqs.~(\ref{eq:EOMbacterium}) with $U_b=0$ in terms of the particle radius $R$ and timescale $R/U$. However, while the organism size $l_\tn{b}$ does not directly affect the dynamics, it determines the interception criterion: we take the sinking particle to be a perfect absorber and stop simulations if any part of the rod or disk touches the particle~(Appendix~\ref{appsec:sim_methods}). We now systematically vary $\ga$ and $l_\tn{b}/R$ and measure the encounter probability and typical interception location.

\begin{figure*}[t!]
  \caption{
Experiments with non-motile elongated diatom cells (\textit{Phaeodactylum tricornutum}) are consistent with the predictions of the model~[Eq.~(\ref{eq:EOMbacterium})] that rods maintain tangential  orientation as they are advected around a sinking particle~(Section~\ref{sec:non-motile_bac}).
(a)~Minimum intensity projection obtained by phase contrast microscopy shows the streamlines of the suspended diatoms around the alginate particle~(Movie~3).
(b)~The experimental ensemble of diatom positions (dots) and the sine of the angle the diatoms make with the flow direction~(see colorbar).
(c)~The corresponding ensemble obtained from simulations, in which we mimic the same information loss as in the experiment: the plot represents rods lying in the focal plane and we dismiss rods with a significant out-of plane component, see text for more details. As predicted in Section~\ref{sec:Stokesflow_asymp_critical}, we observe that rods approaching the particle near the upstream stagnation line orient tangentially to the particle~[yellow region below the particle in (b,c)], while rods that leave the particle near the downstream stagnation line point nearly radially~(blue region above the particle). As they travel near the particle surface, rods tend to maintain tangential orientation but also occasionally tumble.
}
  \centering
    \includegraphics[width=1.0\textwidth]{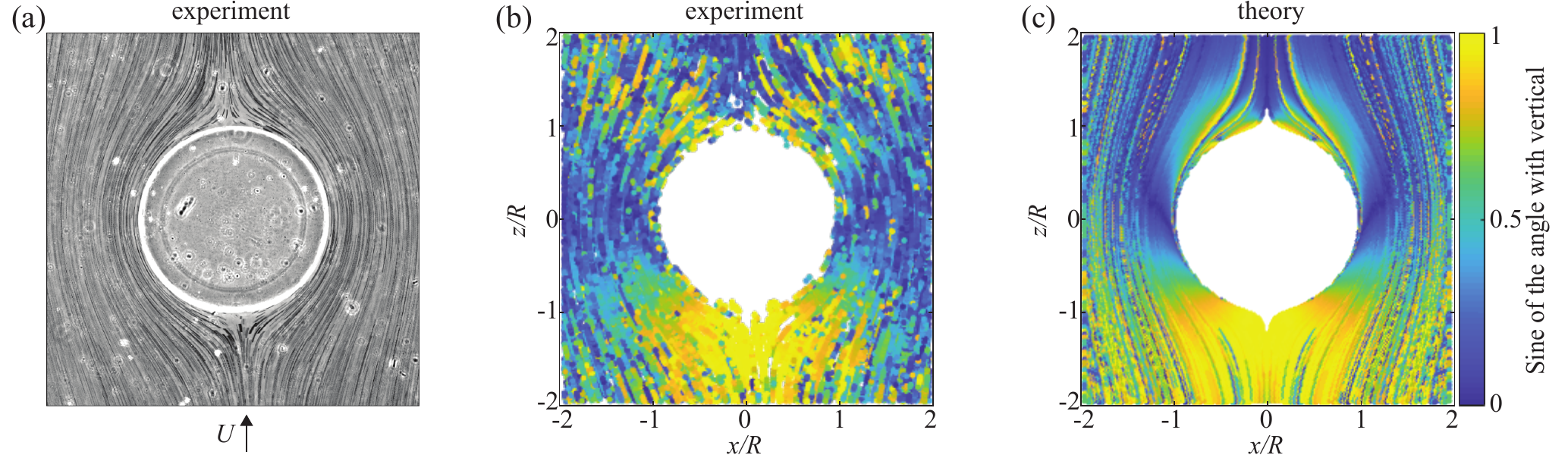}
\label{fig:experiment}
\end{figure*}

The impact of the microorganism aspect ratio $\ga$ and its relative size $l_\tn{b}/R$ on the encounter problem is summarized in Figs.~\ref{fig:nonmotile}(b--d); $l_\tn{b}$ denotes the longer dimension - length for elongated organisms, width for oblate ones. Varying $\ga$~[Fig.~\ref{fig:nonmotile}(b)] at fixed $R/l_\tn{b}=10$, shows that the collision radius of elongated microorganisms is determined by their width, not length~(Movie 1). This is evident from the variation of $P(\rho)$, the interception probability for an initial position at distance $\rho$ from the centerline: $P(\rho)$ decreases sharply from one once $\rho>r_\tn{c}/\alpha$. The probability tail between $r_\tn{c}/\alpha<\rho<r_\tn{c}$ indicates that rods occasionally reorient and use their length to intercept the particle~(Movie 2). However, shear largely suppresses this effect, see the light blue broken line in Fig.~\ref{fig:nonmotile}(b), which represents trajectories without the shear-induced reorientation (parallel transport). In contrast to elongated microorganisms, oblate organisms (disks) utilize their full size to intercept the particle - $P(\rho)$ drops sharply from one to zero near $r_\tn{c}$. To see the impact of varying the relative size $R/l_\tn{b}$, we fixed two aspect ratios, $\ga=10$ [Fig.~\ref{fig:nonmotile}(c)] and $\ga=0.1$ [Fig.~\ref{fig:nonmotile}(d)] and computed $P(\rho)$ as well as the distribution $\xi(\theta)$ for $R/l_\tn{b}=10,100$. We observe that, as the colloid gets smaller (or the sinking particle gets larger), the effects described above become more pronounced, in the sense that for rods the probability tail between $r_\tn{c}/\ga<\rho<r_\tn{c}$ shrinks, whereas for disks $P(\rho)$ approaches a step function with jump at $r_\tn{c}$. Therefore, the formula~(\ref{eq:eta_spheres}) for the encounter efficiency by spherical colloids is replaced by
\be
\label{eq:eta_rods_disks}
\eta_\tn{rods}=[r_\tn{c}/(\ga R)]^2=\eta_\tn{spheres}/\ga^2, \quad \eta_\tn{disks}=\eta_\tn{spheres}
\ee
in the case of (small $R/l_\tn{b}\gg 10$) nonspherical colloids, where rods means $\ga \gg1$ and disks $\ga \ll 1$.

Different interception efficiency for rods but not disks as compared to spherical colloids is consistent with the analytical arguments presented in Section~\ref{sec:Stokesflow_asymp}. Initially, shear aligns rods and disks tangentially to the sinking particle surface as they approach it along the stagnation line~[regions I in Figs.~\ref{fig:particles_fixed_StokesFlow}(b,c)]. As they slide near the particle, both rods and disks experience shear that tries to reorient them away from the tangential configuration, potentially increasing their chance to intercept the particle~[region II in Fig.~\ref{fig:particles_fixed_StokesFlow}(b) for rods and regions II and IIIa in Fig.~\ref{fig:particles_fixed_StokesFlow}(c) for disks]. However, disks are exposed to this reorienting effect over a larger region than rods, suggesting that disks complete this reorientation while near the particle, whereas rods orient radially only when they are too far behind the particle (Movie 1). Occasional interception by rods caused by the reorientation in region II is responsible for the small probability tail in $P(\rho)$ in Figs.~\ref{fig:nonmotile}(b,c)~(Movie 2). In summary, simulations confirm the intuition based on analytical arguments: the collision cross-section for rods is determined by their shorter dimension whereas the opposite is true for disks.

\begin{figure*}[t!]
  \caption{
The shear-shape coupling significantly impacts the encounter rate between motile bacteria and sinking particles and the typical interception location on the particle. Encounter rate kernel $\dot N/n$ (a), encounter efficiency~$\eta$ (b) and mean interception colatitude~$\langle \theta\rangle$ (c) as a function of the sinking speed relative to the bacterial swimming speed $U/U_\tn{b}$ for different bacterial aspect ratios~$\ga$. The continuous lines represent the ballistic model~(\ref{eq:EOMbacterium}) while the broken lines denote the quasi-ballistic model with rotational diffusion~(\ref{eq:EOMdiff}). In the ballistic case, motility, elongation and shear enhance the encounter rate about twofold for slowly sinking particles as compared to the case with shear-induced reorientation switched off~(purple and green lines). However, for intermediate to fast sinking particles, the encounter rate falls orders of magnitude below the value set by the interception of non-motile rods. On the particle, elongated motile bacteria attach preferentially to its leeward side.
}
  \centering
    \includegraphics[width=1.0\textwidth]{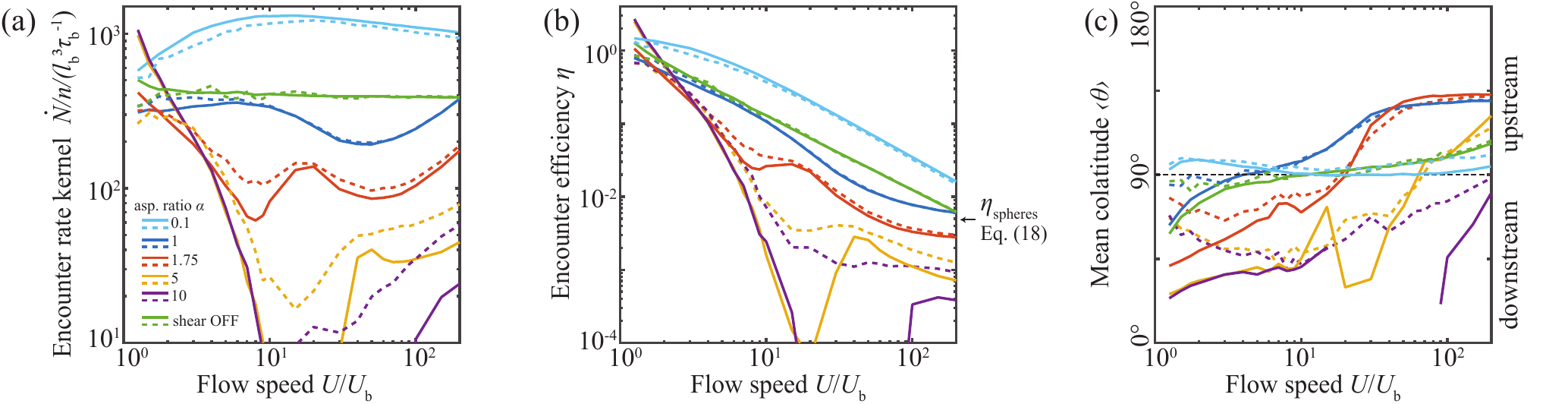}
\label{fig:EC_global}
\end{figure*}

\subsection{Experiments with elongated diatom cells}
\label{sec:experiments}
Selected experiments with non-motile diatom cells confirm the predictions of the model~(\ref{eq:EOMbacterium}) discussed in Sections~\ref{sec:non-motile_bac},~\ref{sec:Stokesflow_asymp_critical} in the case of the non-motile rods~(Fig.~\ref{fig:experiment} and~Movie~3). We ran a suspension of the non-motile, elongated diatom cells \textit{Phaeodactylum tricornutum}~(strain CCMP2561) in sea water at a mean flow velocity of~$\SI{168}{\micro\meter\per\second}$ through a microfluidic channel with a calcium-alginate spherical particle held fixed in the middle by the channel walls~[Fig.~\ref{fig:experiment}(a)]; see Appendix~\ref{appsec:exp_methods} for more details on the experimental protocol. The particle size was $R=\SI{566}{\micro\meter}$, the average diatom length $l_\tn{b}= \SI{21.2}{\micro\meter}$ and their average aspect ratio $\ga=6.8$. Rather than directly estimating the encounter rate, which proved difficult due to the challenge of imaging in the immediate vicinity of the particle, we used image analysis to quantify the orientation that the diatoms assume in the vicinity of the particle in the channel mid-plane. Using this approach, we extracted from 50 consecutive frames an ensemble of diatom positions and orientations~[Fig.~\ref{fig:experiment}(b)]. Since the shear-induced reorientation effects are strongest near the particle~[Fig.~\ref{fig:particles_fixed_StokesFlow}(d)], we focus on diatoms that are at distance  $R<r<2R$ away from the particle center. Importantly, since we only image the focal plane, this ensemble is skewed towards diatoms moving in the focal plane and also oriented in that plane. For this reason, to compare the experimentally determined orientations with those predicted by the model~(\ref{eq:EOMbacterium}), we run additional numerical simulations to mimic the same information loss as in the experiments. Specifically, we simulate a front of uniformly distributed and randomly oriented elongated non-motile rods (with $R/l_\tn{b}=21.2$ and $\ga=6.8$), as in the previous section. We focus on trajectories lying in the particle mid-plane~(as in the imaged region) and extract the rod orientations at positions along the streamlines corresponding to equal time intervals. We reject orientations that have the out-of-plane component larger than $\sin(\ang{30})=0.5$, to mimic the information loss of diatoms that point out of the focal plane in experiments; the results are robust against variation in this threshold~(Fig.~\ref{fig:exp_theory_thresholds}).

The experimentally determined orientations of diatoms agree very well with the numerical results for elongated ellipsoids with the same geometrical characteristics ~[Figs.~\ref{fig:experiment}(b,c)]. In particular, as predicted in Appendix~\ref{sec:Stokesflow_asymp_critical}, diatoms approaching the particle near the upstream stagnation line orient tangentially to the particle surface, while diatoms departing from the particle near the downstream stagnation line point nearly radially [yellow vs. blue regions in~Figs.~\ref{fig:experiment}(b,c)]. Close to the particle surface, diatoms tend to maintain tangential orientation but can also occasionally tumble. Tumbling happens most often when diatoms are to the leeward side of the particle, which is consistent with the action of shear depicted in region II in Fig.~\ref{fig:particles_fixed_StokesFlow}(b), where vorticity dominates over straining and tries to spin rods in the plane of the picture. In summary, these experiments validate detailed aspects of our model for non-motile microorganisms and demonstrate that the effect of shear---shape coupling can be substantial for realistic marine microorganisms.

\begin{figure*}[t!]
  \caption{Motile elongated bacteria preferentially attach to the leeward side of a sinking particle due to hydrodynamic screening and focusing upstream and downstream of the particle. Encounter probabilities $P(\bs x)$ as a function of the position $\bs x$~[Eq.~(\ref{eq:enc_prob_x})] for perfect spheres (a), moderately elongated swimmers (b) and perfect rods (c). The sinking to swimming speed ratio is fixed at $U/U_\tn{b}=3$. (d--f)~Histograms of the interception colatitudes $\theta$ for initial positions in the whole domain shown in (a--c) show a transition from a nearly uniform coverage of the particle by spherical swimmers (d) to preferential leeward attachment for motile rods (f).
(g-i) Representative swimming trajectories (left) and successful initial orientations~(right) for rod-like bacteria starting from the initial positions indicated in (c) illustrate the hydrodynamic screening~(g) and focusing~(h,i). 
}
  \centering
    \includegraphics[width=1.0\textwidth]{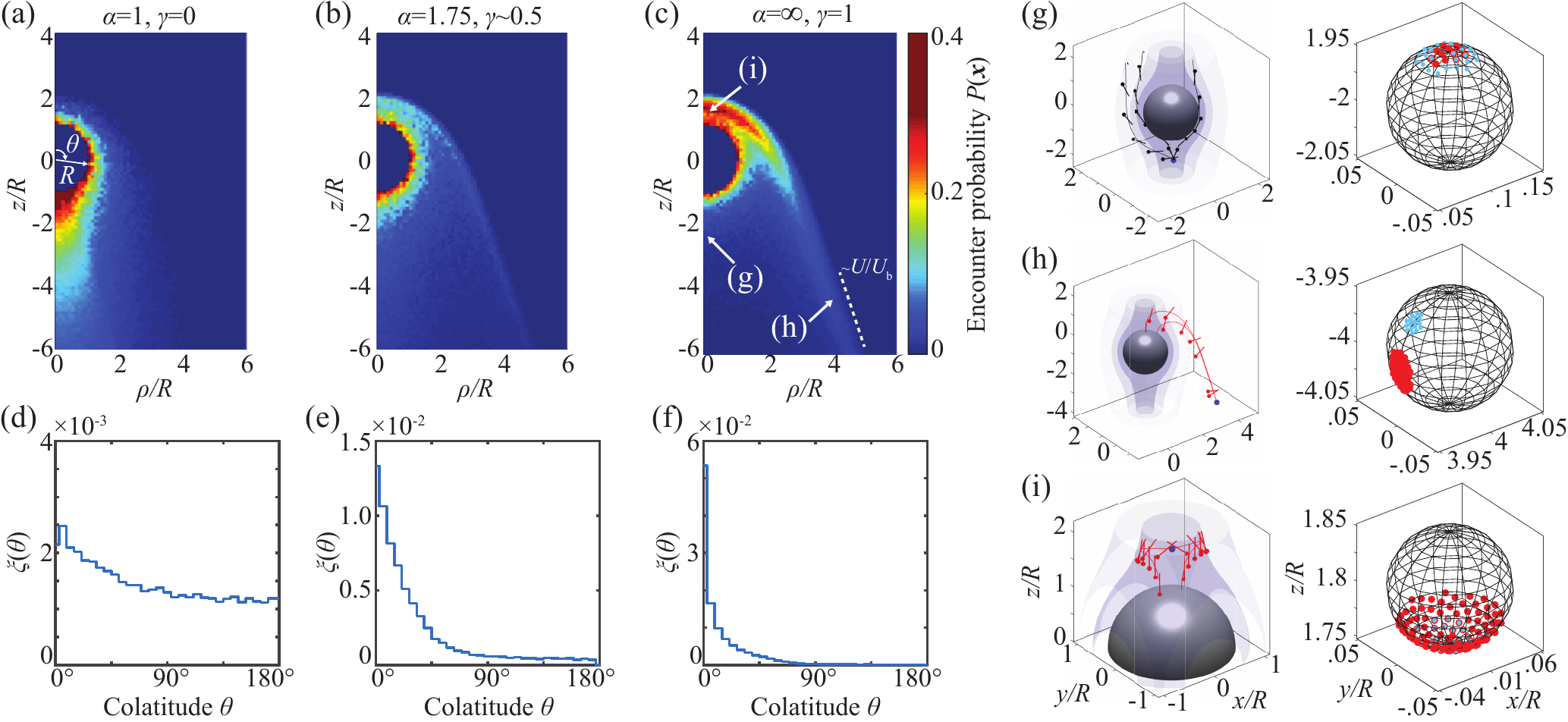}
\label{fig:encounter_prob_heatmap}
\end{figure*}

\section{Motile bacteria}
\label{sec:motile}
The encounter rate between motile elongated bacteria and sinking particles in the ballistic regime depends strongly on the particle sinking speed relative to the bacterial swimming speed~(Section~\ref{sec:motile_bac_er}). For slow sinking particles, shear increases the encounter rate more than twofold and leads to preferential attachment of bacteria to the leeward side of the particle. However, as the sinking speed increases, shear decreases the encounter rate, orders of magnitude below the rate of non-motile organisms. These mechanisms of hydrodynamic focusing and screening are rationalized at the level of individual bacterial trajectories~(Section~\ref{sec:focus_screen}) in terms of the quasi static picture derived in Section~\ref{sec:analytical}. Finally, to connect with the diffusive description of the encounter process, we introduce rotational diffusion to quantify how various stochastic mechanisms, such as Brownian motion or run-and-tumble reorientation, influence the above ballistic description~(Section~\ref{sec:rot_diff}).

\subsection{Encounter rates for motile bacteria}
\label{sec:motile_bac_er}

Nondimensionalization of the ballistic model~(\ref{eq:EOMbacterium}) in terms of the particle radius $R$ and the time scale $R/U$ derived from the sinking speed $U$ shows that the only two dynamically relevant variables are the ratio of the sinking to swimming speeds $U/U_b$ and the bacterial aspect ratio $\ga$. The bacterial size $l_b$ and particle size $R$ enter the problem through the interception condition but otherwise they do not affect the bacterial trajectories, except for the time it takes to execute them. We assume the particle is a perfect absorber and stop the simulations either if any part of the bacterium touches the particle (interception) or the bacterium ends up far behind the particle. In this section, we fix $R/l_b=10$, scan velocities in the range $U/U_b>1$ (sinking speed greater than swimming speed) and consider several aspect ratios $\alpha$~(Fig.~\ref{fig:EC_global}).

For hypothetical spherical or oblate motile bacteria~\cite{Hess2019}, the encounter rate kernel $\dot N/n$ depends weakly on the sinking velocity $U/U_b$ and the encounter efficiency $\eta$ decays monotonically with $U/U_b$; for spherical swimmers, $\eta$ is close to the values obtained with the reorientation by shear switched off~[dark blue and green lines in Figs.~\ref{fig:EC_global}(a,b)]. However, for elongated swimmers, $\eta$ varies strongly with $U/U_b$~[Fig.~\ref{fig:EC_global}(a,b)]. For slowly sinking particles ($1<U/U_b<2$), $\eta\sim 2-3$, implying that the particle collects bacteria from the volume of water two-three times bigger than the geometric cylinder the particle swipes as it sinks. Furthermore, elongated bacteria intercept the particle to the leeward side~[Fig.~\ref{fig:EC_global}(c)]. Interestingly, as the sinking speed increases, the encounter rates drop very rapidly: in the the velocity window $10<U/U_b<100$, the encounter rate of elongated swimmers ($\ga\geq5$) can be orders of magnitude below the value set by the non-motile rods. We next rationalize these encounter rate enhancement and decrease using the concepts of hydrodynamic focusing and screening.

\begin{figure*}[t!]
  \caption{
Encounter probability $P(\rho)$ (a,b) and distribution of interception colatitudes $\xi(\theta)$ (c,d) for bacteria starting at the plane $z=-6R$ with a random initial orientation for different relative sinking speeds $U/U_\tn{b}$ and aspect ratios $\ga$; (a,c) shows the results of numerical simulations of the ballistic model~(\ref{eq:EOMbacterium}) and (b,d) the stochastic model~(\ref{eq:EOMdiff}). In the parameter range considered, rotational diffusion mainly affects elongated swimmers~(bottom panels), for which it decreases the impact of the hydrodynamics focusing at low sinking speeds but also ameliorate the hydrodynamic screening at higher sinking speeds.
}
  \centering
    \includegraphics[width=1.0\textwidth]{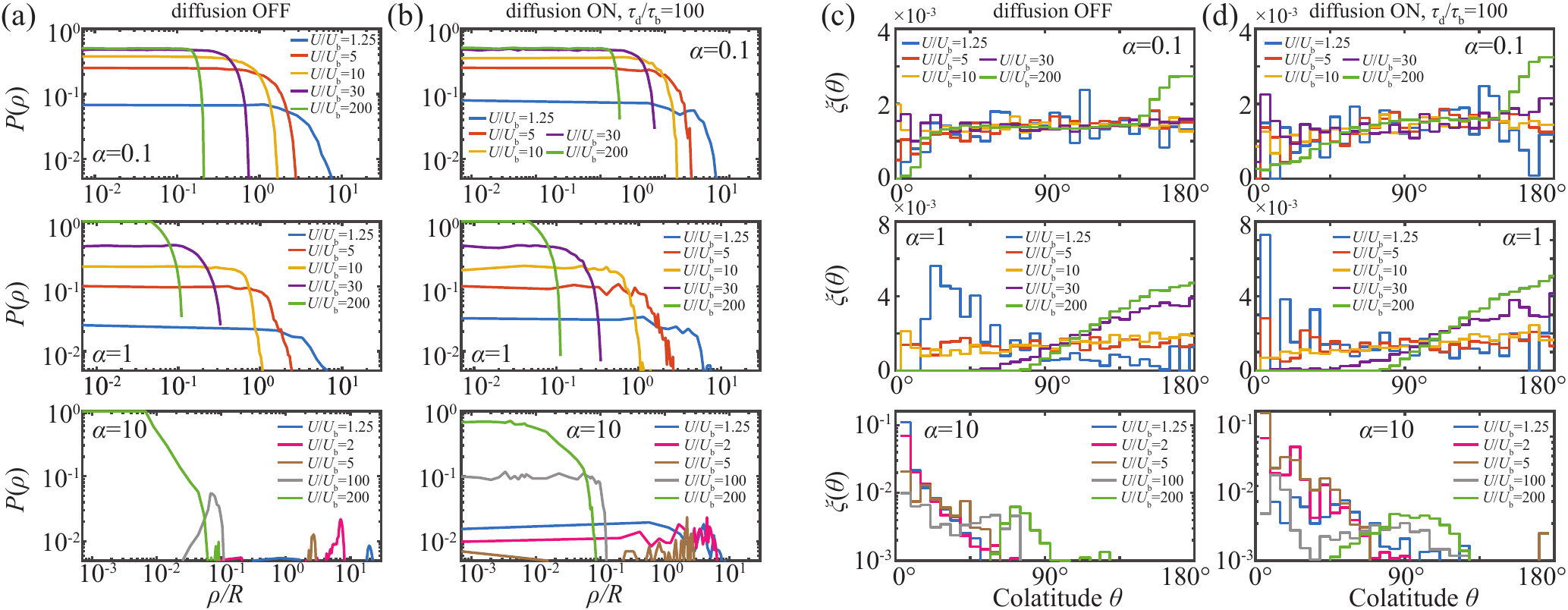}
\label{fig:EC_detailed}
\end{figure*}

\subsection{Hydrodynamic focusing and screening}
\label{sec:focus_screen}
The strong dependence of the encounter efficiency on the particle sinking speed for elongated bacteria~[Fig.~\ref{fig:EC_global}(b)] is a consequence of hydrodynamic focusing and screening. These phenomena are illustrated in Fig.~\ref{fig:encounter_prob_heatmap}, where we compare three swimmers with different aspect ratios; the sinking speed is fixed at $U/U_\tn{b}=3$. Figs.~\ref{fig:encounter_prob_heatmap}(a--c) show the encounter probabilities $P(\bs x)$ for a bacterium starting at $\bs x$ anywhere inside the indicated domain (not just the plane $z=-6R$) with head pointing in a random direction~[Eq.~(\ref{eq:enc_prob_x})]. Since $U/U_\tn{b}>1$, in all three cases there is a cone-like surface of revolution that separates the accessible [$P(\rho)>0$] and inaccessible initial positions - if the bacteria start too far away, they cannot reach the particle. The distribution of $P(\bs x)$ inside the accessibility region vary strongly with $\alpha$. For spheres, $P(\bs x)$ is concentrated below the particle, near the $\rho=0$ stagnation line and decays monotonically to zero with $\rho$ reaching the accessibility horizon~[Fig.~\ref{fig:encounter_prob_heatmap}(a)]. For somewhat elongated swimmers $\alpha=1.75$, the initial positions below the particle become less likely to result in an interception and $P(\bs x)$ starts to concentrate near the edge of the accessible region, which also reaches further out~[Fig.~\ref{fig:encounter_prob_heatmap}(b)]. For perfect swimmers $\alpha=\infty$, the region $\rho\approx0$ is now almost entirely shielded, with $P(\bs x)$ exhibiting a clear high-probability belt at the edge of the accessible region~[Fig.~\ref{fig:encounter_prob_heatmap}(c)]. Far below the particle, the belt slope approaches $\sim U/U_\tn{b}$. Considering the distribution of the interception locations $\xi(\theta)$ for initial positions anywhere in the domains shown in Figs.~\ref{fig:encounter_prob_heatmap}(a--c), elongated swimmers show preferential leeward attachment, with the vicinity of the \lq north pole\rq~being the most likely location~[Figs.~\ref{fig:encounter_prob_heatmap}(d--f)]. We now rationalize the shape of the distributions $P(\bs x)$ and $\xi(\theta)$ at the level of individual swimming trajectories by evoking the quasi-static picture discussed in Section~\ref{sec:Stokesflow_asymp} and shown in Fig.~\ref{fig:particles_fixed_StokesFlow}.

At the level of individual swimming trajectories, the probability~$P(\bs x)$ for spherical swimmers~[Fig.~\ref{fig:encounter_prob_heatmap}(a)] is realized by trajectories that correspond to swimmers initially located below the particle and pointing upwards. However, this intuitive strategy is not available for elongated swimmers because of hydrodynamic screening~[Fig.~\ref{fig:encounter_prob_heatmap}(b,c)]. Recall that, below the particle, shear tends to align rods along the azimuth~[region I in Fig.~\ref{fig:particles_fixed_StokesFlow}(b)]. This shear-induced reorientation coupled with forward motility implies that rod-like swimmers get reoriented and swim away as they approach the sinking particle from below~[Fig.~\ref{fig:encounter_prob_heatmap}(g), Movie 4]. For the same reason, it is very unlikely that elongated swimmers attach to the front of the particle, which explains the small values of $\xi(\theta)$ for colatitudes $\theta>\ang{90}$~[Fig.~\ref{fig:encounter_prob_heatmap}(f)]. Instead, successful interceptions for elongated bacteria must follow a different strategy~[Figs.~\ref{fig:encounter_prob_heatmap}(h,i)]. To avoid the screening, elongated swimmers must start on the belt far away from the centerline of the sinking particle, on the edge of the accessibility horizon. Furthermore, their initial orientations have the be roughly horizontal, pointing towards the centerline~[Figs.~\ref{fig:encounter_prob_heatmap}(h)]. Such initial conditions allow the bacteria to avoid the screening region I of Fig.~\ref{fig:particles_fixed_StokesFlow}(b) and explore the shear-induced radial reorientation in region III. This hydrodynamic focusing then leads to preferential leeward attachment~(Movie~5).

The mechanisms of hydrodynamic focusing and screening described above rationalize the strong dependence of the encounter efficiency $\eta$ on the particle sinking speed $U/U_\tn{b}$ presented in~[Fig.~\ref{fig:EC_global}(b)]. For slowly sinking speeds, both mechanisms are present. However, the high probability belt at the edge of the accessibility horizon for elongated swimmers extends a large volume and hence many swimmers can utilize the focusing effect, which explains why $\eta>1$ in that flow range. However, as $U/U_\tn{b}$ increases, the high probability belt moves closer to the center line since its diameter scales as $U_\tn{b}/U$. This reduces the accessible volume of water at the rate at least $\sim U^{-2}$. Furthermore, as the belt shrinks in diameter, it enters the region of hydrodynamic screening and eventually disappears~[Fig.~\ref{fig:EC_detailed}(a)]. Thus, in the range $10<U/U_\tn{b}<100$, only screening persists, which explains the very small values of $\eta$ in that range. Only for swimming speeds $U/U_\tn{b}>100$, $\eta$ rises again, until it starts to recover the limit set by the interception rate of non-motile rods.

\subsection{Impact of rotational diffusion}
\label{sec:rot_diff}
In the purely ballistic picture of the encounter process outlined in the two previous sections, shear is the only factor responsible for microorganism reorientation. In reality, bacteria experience Brownian rotational diffusion as well as perform run-and-tumble or run-and-reverse dynamics. The combination of these stochastic mechanisms likely interferes with the shear-induced reorientation in a complex manner. As a first step to systematically study the impact of these additional mechanisms, we introduce a single rotational diffusion term to Eq.~(\ref{eq:EOMbacterium})
\bse
\label{eq:EOMdiff}
\be
\label{eq:EOMdiffA}
\dot{\bs x}&=&U_\tn{b}\bs p+\bs v, \\
\label{eq:EOMdiiffB}
\dot{\bs p}&=&(\bs I-\bs p \bs p^\tn{T})[(\gamma E+W)\bs p+\sqrt{2D_\tn{r}}\bs \xi],
\ee
\ese
where $D_\tn{r}$ is the cell's effective rotational diffusivity and $\bs \xi$ is a delta-correlated 3D white noise with zero mean. We express the diffusive timescale $\tau_\tn{d}=D_\tn{r}^{-1}$ in terms of the time $\tau_\tn{b}=l_\tn{b}/U_\tn{b}$ needed for a bacterium to travel the distance equal to its bodylength. 

To study the impact of rotational diffusion on the encounter rates and attachment location, we fixed the diffusive timescale at $\tau_\tn{d}/\tau_\tn{b}=100$, which corresponds to the typical time-scale set by the run-and-tumble motility~\cite{Kiorboe2002}. We then repeated the scans according to the same protocol as in Section~\ref{sec:motile_bac_er}, with $R/l_\tn{b}=10$ and $U/U_\tn{b}>1$~(the broken lines in Fig.~\ref{fig:EC_global}). We find that diffusion has little effect on the encounter rates and attachment location for spherical swimmers. However, for elongated swimmers, diffusion decreases the impact of hydrodynamics focusing at low sinking speeds but also decreases the hydrodynamic screening at higher sinking speeds.

The uniformizing impact of diffusion is studied in more detail in Fig.~\ref{fig:EC_detailed}, where we consider the interception probability $P(\rho)$ for a bacterium starting at the plane $z=-6R$ with random orientation~[Figs.~\ref{fig:EC_detailed}(a,b)] as well as the corresponding distribution of the interception locations $\xi(\theta)$~[Figs.~\ref{fig:EC_detailed}(c,d)]. We compare side to side the cases without~[Figs.~\ref{fig:EC_detailed}(a,c)] and with diffusion~[Figs.~\ref{fig:EC_detailed}(b,d)] for oblate, spherical and elongated swimmers (top, middle and bottom rows, respectively). The accessibility region [defined as $P(\rho)>0$] shrinks under diffusion, because the now erratic motion of bacteria takes longer to reach the particle. and therefore, the bacteria must start closer to the particle to be able to catch it. Within the accessibility region, the distribution $P(\rho)$ for oblate and spherical swimmers is nearly unaffected by diffusion~[Figs.~\ref{fig:EC_detailed}(a,b), top and middle rows] and so is $\xi(\theta)$~[Figs.~\ref{fig:EC_detailed}(c,d), top and middle rows]. However, for elongated swimmers, diffusion decreases the size of the high probability belt near the edge of the accessibility region but also raises the probability of interception for initial conditions directly below the sinking particle~[Figs.~\ref{fig:EC_detailed}(a,b), bottom row]. As a consequence, diffusive elongated swimmers have nonnegligible probability of attaching to the front of the sinking particle~[Figs.~\ref{fig:EC_detailed}(c,d), bottom rows].

\section{Discussion}
\label{sec:discussion}

\begin{figure}[t!]
  \caption{
Given a fixed volume $V_\tn{b}$ of a non-motile microorganism, what shape minimizes or maximizes the encounter efficiency $\eta$ with a sinking spherical particle of volume $V$? Plotting $\eta$ vs the volumetric ratio $V_\tn{b}/V$ reveals that elongation reduces the encounter rate with large particles~($V_\tn{b}/V\to 0$), making rods the optimal shape for avoiding sinking particles, at least as long as rotational diffusion is negligible. Conversely, flattening makes a non-motile microorganism particularly efficient at intersecting the particle. This different behavior of rods and disks is a consequence of fluid shear, which aligns rods tangentially to the sinking particle surface with rare tumbling events, but induces frequent tumbling in the orientation of disks. As the sinking particle becomes large, $V_\tn{b}/V\to 0$, the efficiencies for rods and disks approach the exact expression $\eta=1.4^2 [V_\tn{b}/(V\ga)]^{2/3}$, which follows from Eqs.~(\ref{eq:eta_rods_disks}) after assuming rods and disks can be represented as prolate and oblate ellipsoids with aspect ratio~$\ga$. As the volume ratio $V_\tn{b}/V$ grows, the difference between rods and disks decreases due to increasing tumbling of rods.
}
  \centering
    \includegraphics[width=1.0\textwidth]{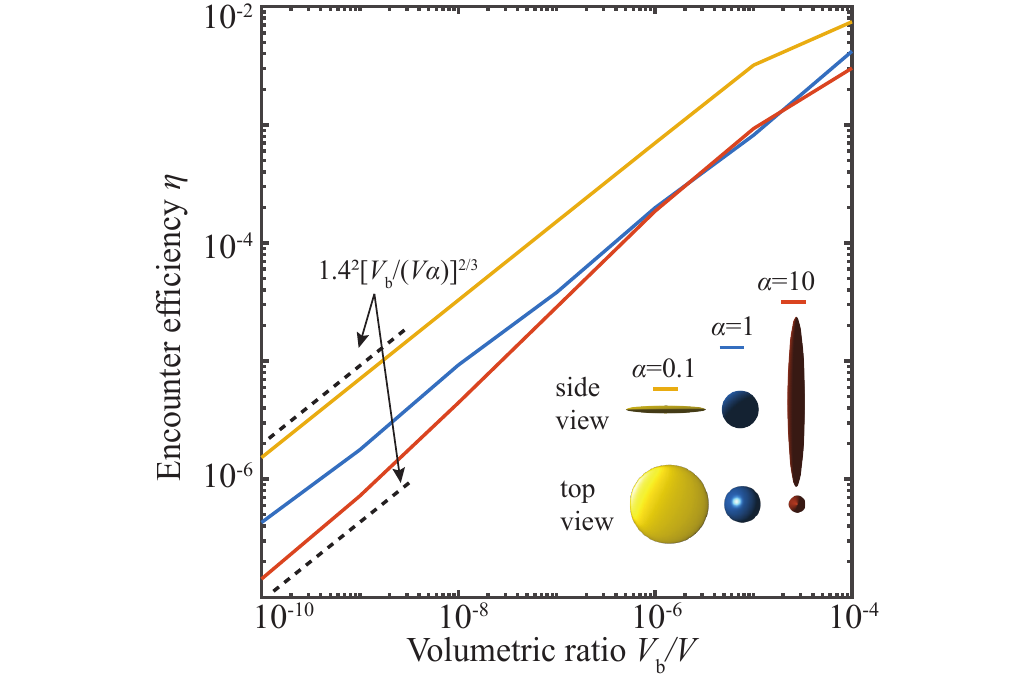}
\label{fig:nonmotile_discussion}
\end{figure}

In this work, we have studied the ballistic limit of the encounter process between microorganisms and sinking particles, with focus on the reorienting effect of shear induced by the sinking particle. Analytical and numerical calculations as well as selected experiments show that the shear---shape coupling acting on a microorganism impacts population-level observables, such as the encounter rate with sinking particles or the typical attachment location on the particle. For the Stokes flow around a spherical sinking particle, rods and disks break the fore-aft symmetry of the flow streamlines, in stark contrast to the behavior of spherical colloids~(Fig.~\ref{fig:particles_fixed_StokesFlow}). This shape-induced symmetry breaking affects the encounter rates~(Figs.~\ref{fig:nonmotile} and~\ref{fig:EC_global}) through mechanisms we have characterized as hydrodynamic focusing and screening~(Fig.~\ref{fig:encounter_prob_heatmap}). Below, we first rephrase these results as solutions to the optimization problem: should a microorganism be elongated or flat, to maximize or minimize the encounter rate with a large moving sphere? Subsequently, we discuss the biophysical consequences of our mechanistic description of the encounter process in the marine environment.

From the perspective of evolution, there are many contexts in which microorganisms may seek to maximize or minimize their encounters with moving objects, including encountering sinking resources~\cite{Kiorboe2002} or symbiotic partners~\cite{Raina2019}, and avoiding predators~\cite{Visser2006}. At the same time these, microorganisms are likely to have other constraints on their volume, such as growth maximization and genome size. For non-motile microorganisms with negligible rotational diffusion~(Section~\ref{sec:non-motile_bac}), we have seen that shear tends to orient rods tangentially to the sinking particle surface as these move around the particle, whereas disk-shaped microorganisms tumble, which makes their longer dimension available for interception~(Section~\ref{sec:non-motile_bac} and Fig.~\ref{fig:nonmotile}). As a consequence, rods have their encounter efficiencies decreased by a factor equal to the square of their aspect ratio compared to spherical colloids with diameter equal to the rod length. Conversely, disks have the same efficiency as spheres with diameter equal to the disk diameter~[Eqs.~(\ref{eq:eta_rods_disks})].

For non-motile bacteria, over a broad range of ratios of cell volume relative to particle volume ($V_\tn{b}/V$), disks are the most efficient shape to intercept a sinking particle, while rods are the least efficient (Fig.~\ref{fig:nonmotile_discussion}). The contribution of the occasional tumbling of rods grows with $V_\tn{b}/V$, decreasing the difference between the efficiencies of rods and disks as cell volume gets larger; rods remain less efficient than disks but catch up with spheres. Note that we cannot increase the volumetric ratio further without violating the approximations used in the Jeffrey equation, since the microorganism size becomes comparable with the sinking particle. Conversely, as the sinking particle grows, $V_\tn{b}/V\to 0$, the encounter efficiencies approach $1.4^2 [V_\tn{b}/(V\ga)]^{2/3}$~(broken lines in Fig.~\ref{fig:nonmotile_discussion}), which follows directly from Eqs.~(\ref{eq:eta_rods_disks}) and reflects the fact that, as the sinking particle grows, rods cease to tumble while disks always tumble~[Fig.~\ref{fig:nonmotile}(c,d)]. In this limiting regime, the efficiency ratio between disks and rods is $(\ga_\tn{rod}/\ga_\tn{disk})^{2/3}$. For example, for rods with $\ga_\tn{rod}=10$ and disks with $\ga_\tn{disk}=0.1$, disks will be about $20$ times more efficient than rods at intercepting large sinking particles.

For motile microorganisms, encounter rates are controlled by hydrodynamic focusing and screening, which develop from the influence of shear on the direction of swimming. Here we still consider the ballistic regime, with negligible rotational diffusion. In addition, the shear---shape coupling for motile cells leads to the sensitive dependence of the encounter rate on the particle sinking speed relative to the bacterial swimming speed~(Section~\ref{sec:motile_bac_er} and Fig.~\ref{fig:EC_global}). Hydrodynamic screening and focusing describe the effect on elongated motile microorganisms of the shear upstream and downstream of the particle, respectively. In hydrodynamic screening, the shear upstream of the particle aligns rods tangentially to the particle surface (as with non-motile cells), resulting in cells swimming away from the particle. For hydrodynamic focusing, rotation from shear in the downstream half of the particle turns the swimming rods towards the particle. For slowly sinking particles, focusing dominates and enhances the encounter rate~(Fig.~\ref{fig:encounter_prob_heatmap}), while at sinking speed significantly higher than the cell swimming speed, screening dominates and the encounter rate drops far below that of non-motile rods. In contrast to elongated organisms, hypothetical spherical and disk-shaped swimmers respond monotonically to changes in the sinking speed~(Fig.~\ref{fig:EC_global}), a consequence of disk-shaped swimmers experiencing hydrodynamic focusing upstream, not downstream, of the particle. For sinking speeds more than twice the swimming speed, this results in disk-shaped microorganisms encountering particles at a higher rate than elongated microorganisms.  

From an evolutionary perspective, this analysis suggests that adopting a disk shape would be optimal for microorganisms (whether non-motile or motile) in order to maximize particle encounter rates at a broad range of sizes and sinking speeds. Of course, this neglects other evolutionary pressures on morphology. However, for particles sinking at speeds close to the organism swimming speeds (i.e., relatively slowly), elongation increases the encounter rates up to twice that of motile disks. Coupled with the large reduction in encounter rates for more rapidly sinking particles, elongation can be viewed as biasing ballistically swimming organisms strongly towards slowly sinking particles. This could be subject to selective pressures for microorganisms in situations where optimal growth occurs near the surface, and therefore rapidly sinking particles are better avoided.   

\begin{figure*}[t!]
  \caption{
To quantify the impact of shear and elongation on the encounter rates and interception locations between bacteria and sinking particles in the ocean, we focus on realistic model parameters and compare three cases: motile elongated bacteria of length $l_\tn{b}=\SI{2}{\micro\meter}$, swimming speed $U_\tn{b}=\SI{50}{\micro\meter\per\second}$ and aspect ratio $\ga=3.3$~(a,d); the same bacteria, but with the shear-induced reorientation switched off~(b,e); and non-motile spherical bacteria of diameter $l_\tn{b}=\SI{1}{\micro\meter}$~(c,f). In these three cases, we compute the encounter efficiency $\eta$ (a--c) and mean interception colatitude $\langle \theta\rangle$~(d--f) as a function of the sinking particle speed $U$ and radius $R$. We consider a range of sinking particle sizes ($R\sim \SI{3}{\micro\meter}-1\tn{mm}$) that covers the most abundant marine particles~\cite{Bochdansky2016}. In the simulations, the rotational diffusion coefficient for motile swimmers was set to $D_\tn{r}=0.25 \tn{s}^{-1}$, while the translational diffusion coefficient of the non-motile spheres was set to $D_\tn{t}=\SI{0.43}{\micro\meter\squared\per\second}$. For such parameters, the panels capture the ballistic and diffusive regimes, as well as the in-between quasi-ballistic regime.
}
  \centering
    \includegraphics[width=1.0\textwidth]{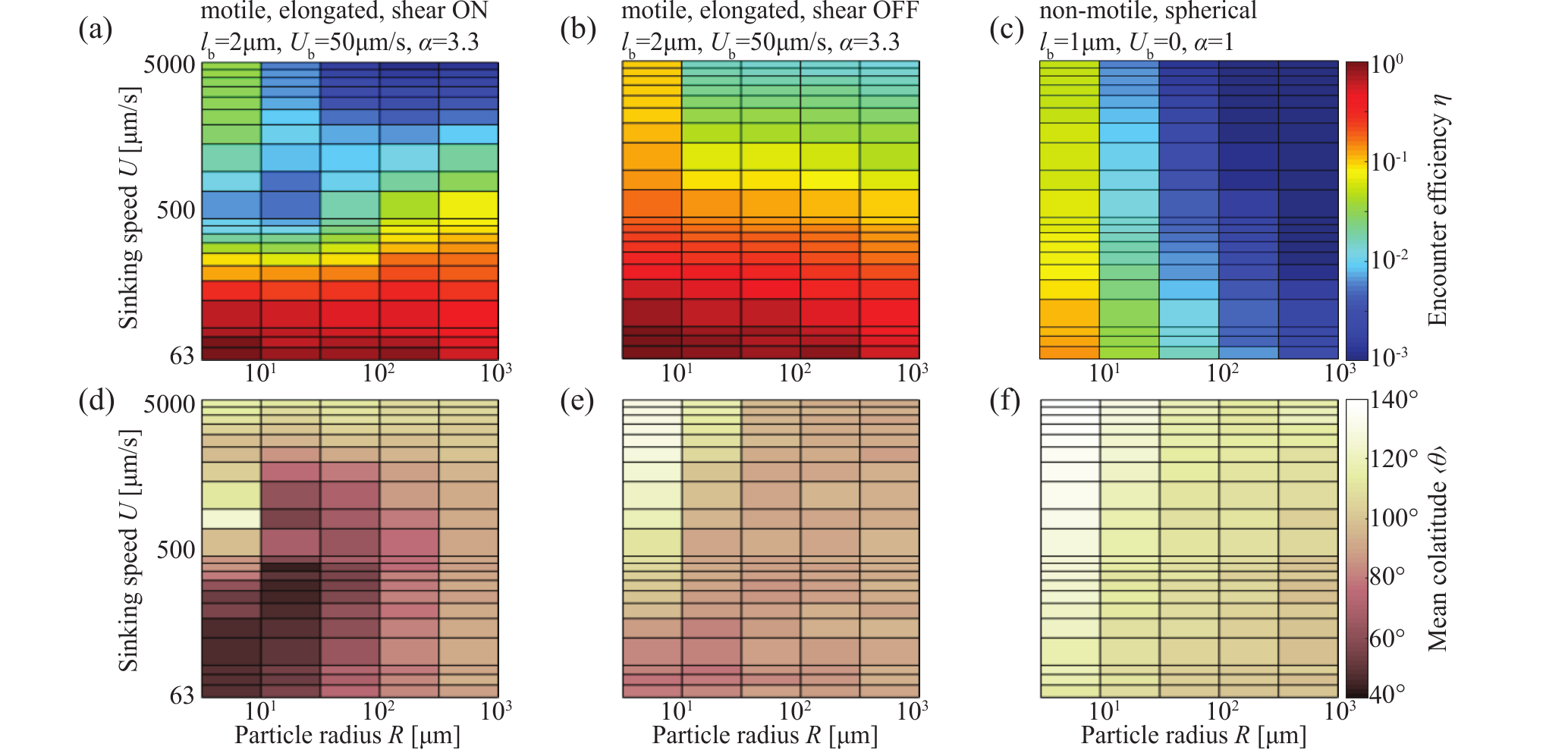}
\label{fig:EC_realistic_bacteria}
\end{figure*}

For the specific case of marine bacteria encountering sinking marine particles, it is necessary to connect the ballistic description of the encounter process with the classical approach based on approximating bacterial motility as a diffusive process~\cite{Karp-Boss1996,Kiorboe2002,Kiorboe2008}. Marine bacteria are subject to various sources of random reorientation, from Brownian rotational diffusion to self-generated run-and-tumble or run-and-reverse motility, where segments of straight swimming are interrupted by randomization of the swimming direction. As a consequence, on scales larger than the bacterial run length and timescales longer than the reorientation time, bacterial motility can be effectively characterized as a diffusive process~\cite{Karp-Boss1996,Kiorboe2002,Kiorboe2008} - this is a general feature of superimposing a large number of uncorrelated random segments~\cite{Chandrasekhar1943}. In this limit, relevant to large sinking particles, the encounter rate is proportional to the bacterial effective diffusion coefficient and the Sherwood number, a flow-induced enhancement factor~\cite{Kiorboe2008}. Additionally, while the shear-induced reorientation in the diffusive limit can be neglected in certain regimes~\cite{Kiorboe2001}, precise quantification of its impact on the encounter rate and attachment locations across a wide range of particle speeds and sizes might require kinetic theory approach~\cite{Chandrasekhar1943,Saintillan2008,Bearon2015}. However, as the particle becomes smaller or the sinking speed increases, the system becomes ballistic and the diffusive approximation overestimates the encounter rate. The reason for this overestimation comes from the fact that the encounter probability for a bacterium at distance $r$ from the particle decays as $r^{-1}$ in the diffusive regime, and as $r^{-2}$ in the ballistic regime~\cite{Berg1977}, at least for stationary particles and without shear. Since hydrodynamic interactions can significantly bend bacterial trajectories~\cite{Rusconi_NatPhys2014,Secchi_2019,Mino2018}, the need to go beyond arguments based on straight-line swimming motivated the above study of the pure ballistic limit. For the marine application, we consider ballistic motile bacteria supplemented by rotational diffusion~(Section~\ref{sec:rot_diff}), which we have shown reduces the strength of hydrodynamic screening on rod-shaped swimming cells.

We consider two major classes~\cite{Kiorboe2002}: motile elongated bacteria and non-motile spherical bacteria. For motile elongated bacteria, we evaluate the predictions of our model for cells of length $l_\tn{b}=\SI{2}{\micro\meter}$, swimming speed $U_\tn{b}=\SI{50}{\micro\meter\per\second}$ and aspect ratio $\ga=3.3$. For non-motile spherical bacteria, we choose a diameter $l_\tn{b}=\SI{1}{\micro\meter}$. These represent typical characteristics of motile copiotrophic bacteria that actively seek and engage marine particles, and more oligotrophic non-motile bacteria which may nevertheless encounter and stick to particles.  For these representative marine bacteria, as well as the equivalent motile bacteria without the influence of shear, the corresponding encounter efficiency $\eta$ [Fig.~\ref{fig:EC_realistic_bacteria}(a--c)] and mean interception colatitude $\langle \theta\rangle$ [Fig.~\ref{fig:EC_realistic_bacteria}(d--f)] have been computed as a function of the sinking particle speed $U$ and particle radius $R$. The range of sinking speeds we consider is $\SI{60}{\micro\meter\per\second}-\SI{5}{\milli\meter\per\second}$. The range of particles sizes, $\SI{3}{\micro\meter}-\SI{1}{\milli\meter}$, covers the most abundant marine sinking particles~\cite{Bochdansky2016}. For elongated bacteria, randomization of orientation is effectively represented by a single rotational diffusion coefficient $D_\tn{r}=\SI{0.25}{\per\second}$; the diffusive timescale~($\sim\SI{4}{\second}$) gives the run length of about $\SI{200}{\micro\meter}$, typical of marine bacteria~\cite{Kiorboe2002}. For such a run length, the range of particle sizes spans the ballistic and diffusive regimes as well as the intermediate transition range. Finally, the translational diffusion of the non-motile spherical bacteria was set to $D_\tn{t}=\SI{0.43}{\micro\meter\squared\per\second}$, which represents Brownian motion of a micron-sized sphere at room temperature~\cite{Bechinger2016}. Under these conditions, accounting for the shear reorientation of motile marine bacteria substantially reduces the encounter efficiency for small fast-sinking particles and alters the location of encounters for slow-sinking particles below 100 microns in radius. In contrast, the diffusive spherical swimmers show weak dependence of encounter efficiency on sinking speed, but much greater sensitivity to particle size.

\begin{figure}[t!]
  \caption{
Comparison of the encounter efficiencies between motile elongated and non-motile spherical microorganisms~(a) and between motile elongated microorganisms with and without shear~(b) computed from the panels (a--c) of Fig.~\ref{fig:EC_realistic_bacteria}. For reference, we plot the absolute values of the sinking/raising speeds as determined by the Stokes law~(\ref{eq:terminal_speed}) for sinking particles with densities~$\rho_\tn{p}$ higher than the density of water~$\rho_\tn{w}$ by $5-15\%$ (in black font), such as the fast sinking fecal pellets, as well as raising bubbles with $\rho_\tn{p}/\rho_\tn{w}\approx 0$ (in white font). The white dashed line denotes the ratio equal to unity.
}
  \centering
    \includegraphics[width=1.0\textwidth]{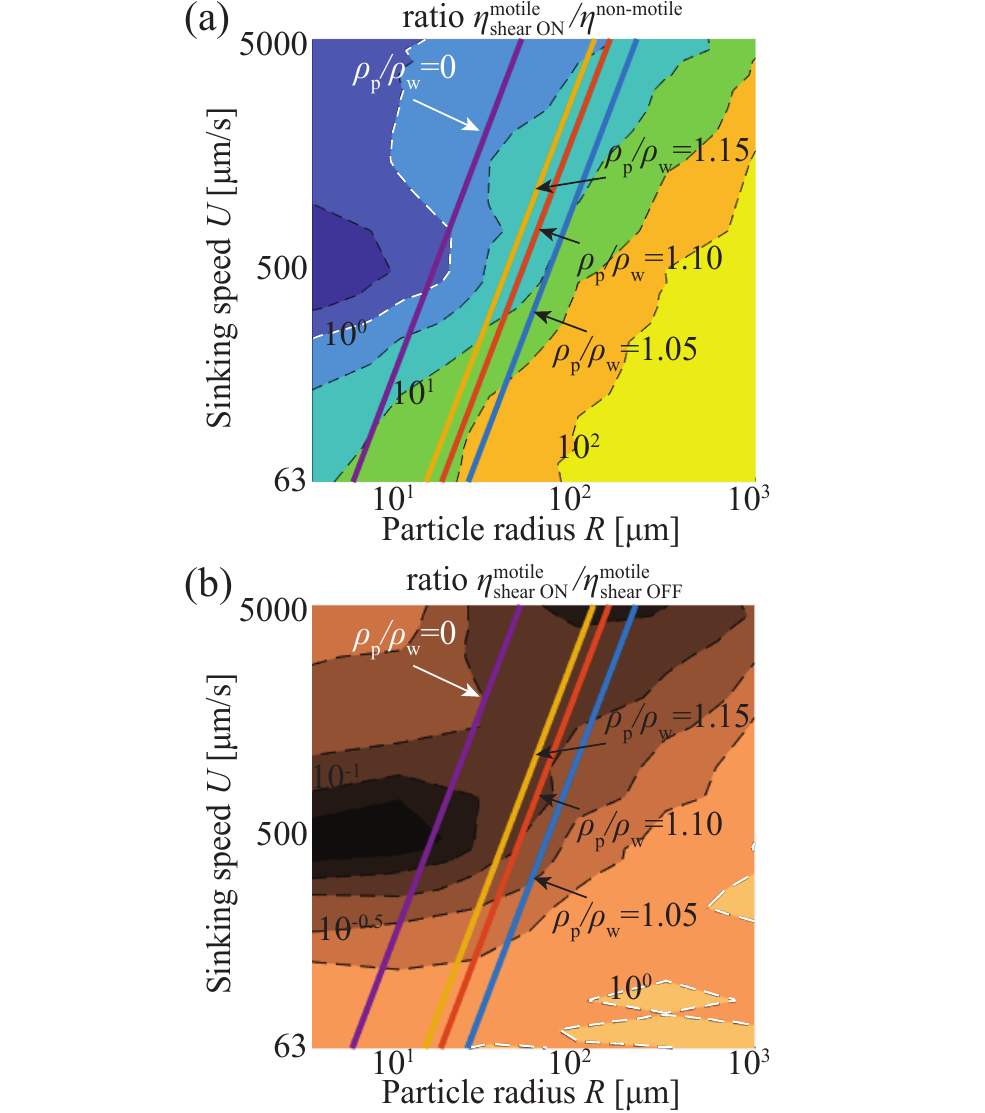}
\label{fig:EC_realistic_bacteria_comparison}
\end{figure}

Since the range of particle sizes considered in Fig.~\ref{fig:EC_realistic_bacteria} captures the ballistic--diffusive transition, the standard computation based on a diffusive analogy only matches the encounter rates of marine bacteria (neglecting shear) for the largest particles with radii approaching $1$ mm~(Fig.~\ref{fig:SI_eta_comparison}). For bacteria with higher rotational diffusivity, this occurs for smaller particle sizes. As the particles get smaller, the diffusion-based calculation starts to overestimate the encounter rate - in the ballistic limit, with particle sizes reaching tens of microns, the two descriptions can differ by more than two orders of magnitude~(Fig.~\ref{fig:SI_eta_comparison}). We now describe in detail the encounter process in the intermediate quasi-ballistic regime, highlighting the role of bacterial motility and fluid shear.

Factoring in shear interactions, motile bacteria  encounter sinking particles at a rate one or two orders of magnitude~[Fig.~\ref{fig:EC_realistic_bacteria_comparison}(a)] greater than non-motile bacteria. This motility-based enhancement factor is smaller by one or two orders of magnitude (depending on particle size) as compared to what would be predicted by the fully diffusive model. The exception when motility decreases the chances of interception ($\eta_\tn{shear ON}^\tn{motile}/\eta^\tn{non-motile}<1$) corresponds to small and very quickly moving objects. This upper-left part of the panel, close to the full ballistic regime, is dominated by hydrodynamic screening and is probably not relevant for marine particles, even for fast sinking-fecal pellets, since their density differs from that of seawater only by about 10\%-20\%~\cite{Bruland1981,Komar1981}, see the red and yellow lines in Fig.~\ref{fig:EC_realistic_bacteria_comparison}, which represent the Stokes law~(\ref{eq:terminal_speed}). However, this hydrodynamic screening regime may be relevant for interception by air bubbles, whose vertical speed is high in view of their large density difference with seawater~\cite{Weber1983}~(purple line in Fig.~\ref{fig:EC_realistic_bacteria_comparison}). Furthermore, comparing the shear on-off cases for motile elongated bacteria~[Fig.~\ref{fig:EC_realistic_bacteria_comparison}(b)], we find that the major impact of shear is to reduce the encounter rates with small particles sinking at intermediate or rapid rates  by up to a factor of~10. This reduction is a consequence of the competition between the hydrodynamic screening of rods upstream of the sinking particle and rotational diffusion. It suggests that elongation-induced screening may be a passive mechanism that allows motile elongated marine microorganisms to prioritize slowly sinking aggregates, at least in the quasi-ballistic particle size range. 

For slowly sinking particles in the quasi-ballistic regime~(lower parts of the panels in Fig.~\ref{fig:EC_realistic_bacteria_comparison}), the observed encounter efficiencies of motile marine bacteria are close to the case without shear, consistent with earlier studies~\cite{Kiorboe2001}. However, at the level of individual trajectories, this encounter rate is realized by hydrodynamic focusing, which results in most bacteria attaching to the leeward side~($\theta<\ang{90}$) of the sinking particle~[compare Figs.~\ref{fig:EC_realistic_bacteria}(d) and~(e); see also Fig.~\ref{fig:SI_cum_sum}]. For small and slowly sinking particles ($R<\SI{50}{\micro\meter}$, $U<\SI{500}{\micro\meter\per\second}$), for which shear dominates over rotational diffusion, more than 75\% of the interceptions occur on the leeward side of the particle. Furthermore, about 25\% of the interceptions are concentrated inside the \lq Arctic circle\rq~($\theta<\ang{23}$), which represents a more than five-fold increase as compared to a uniform coverage of the particle. This leaves the southern hemisphere depleted of bacteria, with almost no interceptions below the~\lq tropic of Capricorn\rq~($\theta>\ang{113}$). Thus, the leeward stagnation point is a flow-induced hotspot where motile and elongated bacteria concentrate due to shear. Since non-motile bacteria intercept the particles on the upstream side~[Fig.~\ref{fig:EC_realistic_bacteria}(f)], in the southern hemisphere, we conclude that flow and shear lead to a bipolar segregation of motile and non-motile marine bacteria on the two sides of a sinking particle.

Although this work has assumed particles to be spherical despite the variety of observed shapes exhibited by marine snow aggregates~\cite{Bochdansky2016}, we expect the phenomena of hydrodynamic focusing and screening of elongated bacteria to be robust to variation in shape. The focusing and screening effects rely on different orientational responses of small rods upstream and downstream of the particle - the key property of the flow that is required for this fore--aft symmetry breaking is the expansion of the streamlines to the front of the particle and their recombination to the back, as well as the no-slip boundary conditions on the particle surface. As long as such general streamline organization is preserved, the effects here described should be robust: while fluid parcels roll on the particle surface~(no slip), they stretch upstream of the particle~(streamline expansion) but compress downstream of the particle~(streamline recombination). This basic process will hold for objects at low Reynolds numbers with no-slip surfaces, and one would therefore expect hydrodynamic focusing and screening of motile elongated bacteria to occur for marine particles in general. 

\section{Conclusions}
\label{sec:conclusions}
In this work, we combined analytical and numerical calculations to estimate the encounter rates between non-motile and motile microorganisms of different morphologies and sinking particles in the ballistic regime relevant for the most abundant small sinking particles. Previous estimates have primarily focused on the diffusive regime, effectively assuming that particles are much larger than the bacterial run length. In the ballistic range, bacterial reorientation becomes a significant factor influencing the encounter process, while it is absent by necessity from diffusive models. We have focused on the coupling between microorganism shape and fluid shear induced by the particle, since shear is the dominant external factor responsible for bacterial reorientation. We have shown that the shape---shear coupling can significantly affect the encounter rate and attachment location on a particle for both non-motile and motile microorganisms. 

For non-motile organisms, shear from a sinking particle can significantly alter the encounter rates of organisms with different morphologies. For elongated organisms, this influence occurs by aligning the cells' long axis tangentially to the particle surface, and was experimentally validated. When the timescale of rotational diffusion is long with respect to particle interactions, shear from sinking particles interacts with the aspect ratio of non-motile organisms to potentially reduce encounter rates by a factor proportional to the square of the aspect ratio. As a result, encounters could exert evolutionary pressure on non-motile cell morphology~\cite{Persat2014}, favoring elongated or disk-like shapes depending on whether encounters are unfavorable or favorable, respectively. 

 For motile microorganisms, interactions with the shear from a sinking particle give rise to two phenomena, hydrodynamic screening and focusing, that alter both the rates and locations of encounters. Elongation helps organisms intercept slowly sinking particles but dramatically reduces the encounter rate with rapidly sinking particles. In contrast to rods, motile disks experience upstream focusing, leading to high efficiency at intercepting rapidly sinking particles. From the perspective of the particle, motile elongated microorganisms typically attach to the leeward side of the particle, while motile disks cover it more uniformly. Under realistic parameters relevant to marine bacteria and sinking particles, which include the effect of randomization of swimming direction from rotational diffusion, hydrodynamic screening leads to a ten-fold decrease in the interception rate of rapidly sinking aggregates, as compared to motility without the shear-induced reorientation. This reduction in encounter rate suggests that elongation-induced screening may be a passive mechanism that allows motile elongated marine microorganisms to avoid rapidly moving particles. Last but not least, motile elongated bacteria attach to the leeward side of the particle, whereas non-motile bacteria attach to the front. Thus, hydrodynamic focusing is a physical source of heterogeneity in particle colonization characterized by bipolar segregation of motile and non-motile microorganisms, which may influence the degradation rate of marine snow aggregates. Whether in terms of encounter rates or encounter locations, these results indicate that the impact of shear reorientation cannot be neglected when evaluating interactions between motile organisms and sinking particles.

The dynamics of shear-driven reorientation are directly relevant to the colonization of marine particles by bacteria. It is well established that motility can greatly enhance the encounter rate of bacteria with sinking particles~\cite{Visser2006,Kiorboe2002}. This enhancement is often estimated via the ratio of the effective diffusivity due to motility and the diffusivity due to Brownian motion, which can be as large as 1000 for highly motile marine bacteria~\cite{Kiorboe2002,Lambert_2019,Berg1993}. The more accurate theory developed here, which accounts explicitly for the interaction between flow and motility in elongated bacteria, refines this estimation in a manner that depends on the particle size and sinking speed relative to the bacterial motility. For particles substantially larger than the bacterial run length, the enhancement in attachment due to motility estimated by the ratio of effective diffusivities and neglecting the impact of shear is increasingly more accurate. For the marine bacteria modeled here, this corresponds to particles with radius greater than approximately $\SI{1}{\milli\meter}$. For smaller particles, which form the bulk of particles in the ocean~\cite{Stemmann2012}, this work reveals that the enhancement in encounters resulting from motility is more moderate and is further reduced as the particle sinking speed increases. In extreme cases, potentially applicable to some bubbles, motility may confer no benefit in encounter rates. However, in the context of marine particles, motility still enhances encounters by one to two orders of magnitude. Since the enhancement in encounter rate due to motility is greater for slowly-sinking particles, this also highlights the potential significance of neutrally buoyant particles~\cite{Mari2017} to motile bacteria. This fundamental knowledge of encounter rates will be a valuable asset in future efforts to rationalize the community composition on marine particles, and ultimately the role of different groups of bacteria in particle degradation and the ocean's biological pump.

In a different domain, the mechanisms of hydrodynamic focusing and screening of rods and disks here described are relevant to the classical filtration problem~\cite{Friedlander_AIChEJ1957}, because our results suggest that shear renders elongated non-motile colloids more difficult to collect than oblate ones. Furthermore, fabrication of Janus-type artificial swimmers makes it possible to build microscale motile objects with different shapes and swimming speed~\cite{Bechinger2016}, and these parameters could be tailored to enhance or suppress the focusing and screening effects. For example, the efficiency in capturing moving spheres may be important in applications such as targeted drug delivery~\cite{Wang2012} and micromachine-enabled decontamination~\cite{Gao2014}. 

In summary, we have demonstrated that hydrodynamic interactions between a small ellipsoid and a large moving sphere break the fore--aft symmetry of the flow streamlines, leading to practical consequences for microorganisms. This symmetry breaking is a consequence of fluid expansion and recombination upstream and downstream of the sphere, but is only revealed when the full tensorial character of the velocity gradient is accounted for, including its straining and rotational components. Such asymmetric two-body couplings are ubiquitous, since they arise when a small nonspherical particle travels near a larger obstacle in a fluid; we have experimentally verified their impact in the case of non-motile elongated diatoms advected around an alginate bead. In the context of swimming bacteria intercepting a sinking particle, hydrodynamic focusing and screening have practical ecological impacts, but applications to other natural or man-made systems are yet to be explored.

\onecolumngrid
\appendix
\counterwithin{figure}{section}

\section{Supplementary Information}
\begin{figure*}[t!]
  \caption{
Additional comparison between experiments with non-motile elongated diatoms~(a) and simulations~(b--d) shown in Fig.~\ref{fig:experiment} of the Main Text. In simulations, we reject rods with out-of-plane components larger than: \ang{15}~(b), \ang{30}~(c) and \ang{45}~(d). Panels (a) and (c) are the same as panels (b,c) in Fig.~\ref{fig:experiment} of the Main Text.
}
  \centering
    \includegraphics[width=1.0\textwidth]{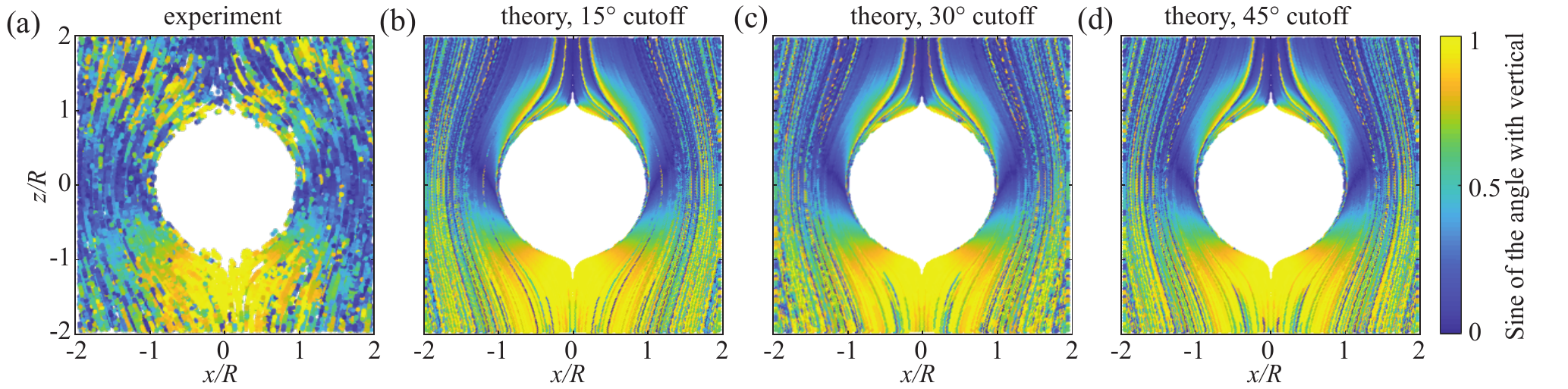}
\label{fig:exp_theory_thresholds}
\end{figure*}

\begin{figure*}[t!]
  \caption{
Ratios of the encounter efficiencies with and without the impact of shear, $\eta_\tn{shear ON}^\tn{motile}$~(a) and $\eta_\tn{shear OFF}^\tn{motile}$~(b), and the encounter efficiency based on the classical diffusive calculation $\eta_\tn{diffusive}^\tn{motile}$; all parameters are the same as in Fig.~\ref{fig:EC_realistic_bacteria} and Fig.~\ref{fig:EC_realistic_bacteria_comparison} of the Main Text. In the presence of flow, the diffusive encounter efficiency is given by $\eta_\tn{diffusive}^\tn{motile}=4 \tn{Sh/Pe}$, where Sh and Pe are the Sherwood and P\'{e}clet number, respectively~\cite{Kiorboe2008}. For the Sherwood number, we used the following formula valid for low Reynolds number $\tn{Sh}=0.5 [1+(1+2 \tn{Pe})^{1/3}]$. For the P\'{e}clet number, we took $\tn{Pe}=U R/D_\tn{b}$, with the bacterial diffusivity $D_\tn{b}=0.5 U_\tn{b}^2 \tau_\tn{d}$, where $\tau_\tn{d}=D_\tn{r}^{-1}$; for the parameters used, $D_\tn{b}=5 \times 10^{-5} \tn{cm}^2/\tn{s}$. As discussed in Section~\ref{sec:discussion}, the diffusive encounter efficiency overestimates the ballistic one and only for the largest sinking particles considered here the two descriptions start to become comparable.
}
  \centering
    \includegraphics[width=0.5\textwidth]{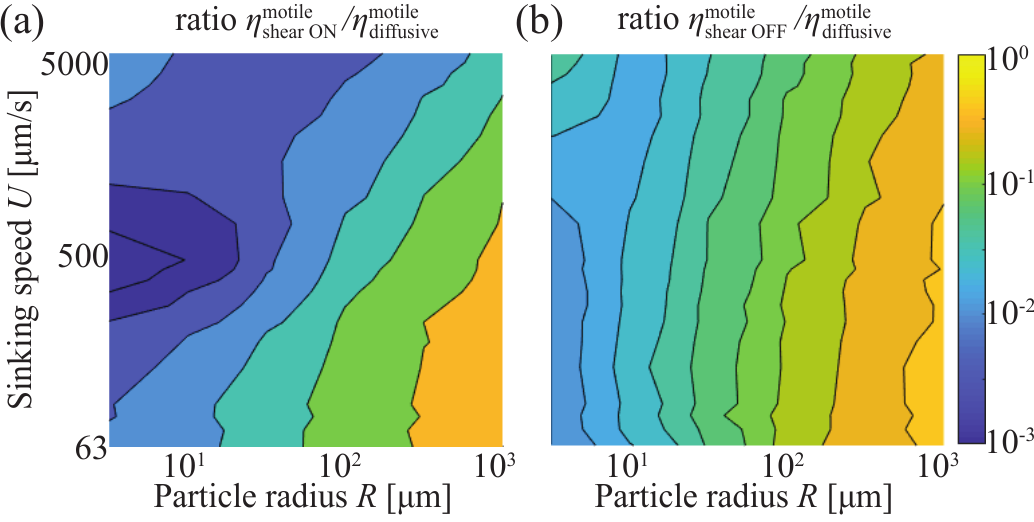}
\label{fig:SI_eta_comparison}
\end{figure*}

\begin{figure*}[t!]
  \caption{
Additional characterization of the landing distribution function $\xi(\theta)$ for the simulations presented in Fig.~\ref{fig:EC_realistic_bacteria} of the Main Text. Here, we look at the colatitude $\theta_f$ such that the fraction $f$ of the interception positions lies in between $\ang{0}<\theta<\theta_f$. Formally, $\theta_f$ is defined as the integral $2\pi\int_0^{\theta_f}\xi(\theta)\sin\theta d\theta=f$. We display the results for three fractions, $f=0.25$~(top row), $f=0.5$~(middle row) and $f=0.75$~(bottom row). For reference, we note that a uniform coverage of the sphere corresponds to $\theta_{0.25}=\ang{60}$, $\theta_{0.5}=\ang{90}$ and $\theta_{0.75}=\ang{120}$.
}
  \centering
    \includegraphics[width=1.0\textwidth]{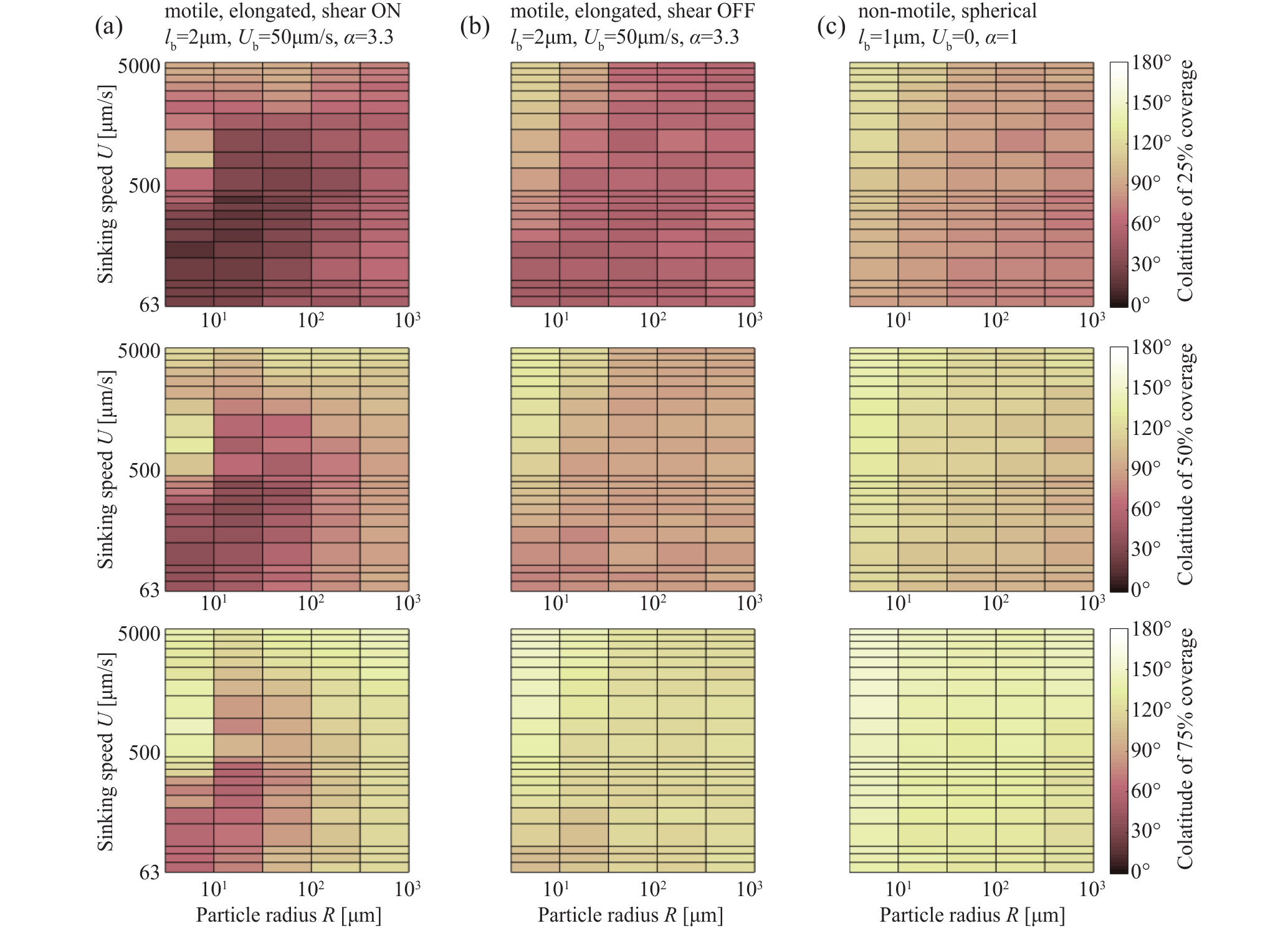}
\label{fig:SI_cum_sum}
\end{figure*}

The Appendix is organized as follows: we provide linear stability analysis of the fixed points of the Jeffery equation in Section~\ref{Appsec:Jeff_linearstability} and derive the limit cycle solutions and their period in Section~\ref{Appsec:Jeff_limitcycle}; these results were discussed in Section~\ref{sec:Jeffrey_asymp} of the Main Text. The velocity gradient for the Stokes flow is derived in Section~\ref{Appsec:AStokesFlow}; its matrix form was used in Eq.~(\ref{eq:VGspherical_ortho}) in Section~\ref{sec:Stokesflow_asymp}~of the Main Text. In Section~\ref{sec:Stokesflow_asymp_critical}, we complement the discussion in Section~\ref{sec:Stokesflow_asymp} of the Main text by analyzing the structure of the velocity gradient on the stagnation lines and the sinking particle surface. The subsequent Sections give details on the numerical simulations~(Section~\ref{appsec:sim_methods}) and experimental methods~(Section~\ref{appsec:exp_methods}). Finally, the three additional figures supplement the Main Text as follows: Fig.~\ref{fig:exp_theory_thresholds} shows that the results presented in Fig.~\ref{fig:experiment} of the Main Text are robust to variation in the cut-off threshold for rejecting the out-of-plane components of rods in the simulations, Fig.~\ref{fig:SI_eta_comparison} quantifies the overestimate in the encounter efficiencies as predicted by the classical diffusive arguments in the range of parameters discussed in Figs.~\ref{fig:EC_realistic_bacteria} and~\ref{fig:EC_realistic_bacteria_comparison} of the Main Text, and Fig.~\ref{fig:SI_cum_sum} provides an additional characterization of the landing distributions for the simulations presented in~Figs~\ref{fig:EC_realistic_bacteria} of the Main Text.

\subsection{Stability analysis of the fixed points of the Jeffrey equation}
\label{Appsec:Jeff_linearstability}
In this section, we analyze the linear stability of the fixed points of the Jeffrey Eq.~(\ref{eq:JeffreysEq2})
\be
\label{eq:JeffreysEq2appen}
\dot{\bs p}&=&(\bs I-\bs p \bs p^\tn{T}){A}^{\gamma}\bs p.
\ee
This analysis will also yield the characteristic timescales of the convergence onto the asymptotically stable solutions. As discussed in Section~\ref{sec:Jeffrey_asymp} and in~\cite{Junk_JMFM2007}, the fixed points of Eq.~(\ref{eq:JeffreysEq2appen}) are given by the real eigenvectors of ${A}^{\gamma}$. Let $\bs \lambda$ be a normalized real eigenvector of ${A}^{\gamma}$ with eigenvalue $\lambda$. Linearizing~Eq.~(\ref{eq:JeffreysEq2appen}) around $\bs\lambda$ by writing $\bs p=\bs\lambda+\Delta\bs p$, where the perturbation $\Delta\bs p$ lies in the tangent space to the sphere at $\bs \lambda$, gives
\be
\label{eq:JeffLinear}
\dot{\Delta\bs p}
&=&
-\lambda \Delta \bs p
+
(\bs I-\bs \lambda \bs \lambda^\tn{T})({A}^{\gamma}\Delta\bs p).
\ee
Eq.~(\ref{eq:JeffLinear}) is a two-dimensional linear dynamical system whose stability can be classified using the standard trace-determinant characterization. To be more explicit, we introduce the basis vectors $\{\bs e_1,\bs e_2\}$ for the tangent space to the sphere at $\bs\lambda$
\be
\bs e_1= \bs n\times \bs\lambda, \quad \bs e_2=\bs e_1\times\bs\lambda,
\ee
where $\bs n$ is an arbitrary nonzero vector, noncolinear with $\bs \lambda$. In that basis, the perturbation reads $\Delta\bs p=\alpha(t) \bs e_1+\beta(t)\bs e_2$ and the linearized system~(\ref{eq:JeffLinear}) reduces to
\be
\label{eq:JeffLinearM}
\begin{bmatrix}
      \dot\alpha  \\[0.3em]
      \dot\beta 
\end{bmatrix},
=
-\lambda
\begin{bmatrix}
      \alpha  \\[0.3em]
      \beta 
\end{bmatrix}+
\begin{bmatrix}
      \bs e_1^\tn{T} {A}^{\gamma} \bs e_1 & \bs e_1^\tn{T} {A}^{\gamma} \bs e_2  \\[0.3em]
      \bs e_2^\tn{T} {A}^{\gamma} \bs e_1 & \bs e_2^\tn{T} {A}^{\gamma} \bs e_2 
\end{bmatrix}
\begin{bmatrix}
      \alpha  \\[0.3em]
      \beta 
\end{bmatrix}=
M_{\bs \lambda}
\begin{bmatrix}
      \alpha  \\[0.3em]
      \beta 
\end{bmatrix},
\ee
where
\be
M_{\bs \lambda}=
\begin{bmatrix}
      \bs e_1^\tn{T} {A}^{\gamma} \bs e_1 -\lambda & \bs e_1^\tn{T} {A}^{\gamma} \bs e_2  \\[0.3em]
      \bs e_2^\tn{T} {A}^{\gamma} \bs e_1 & \bs e_2^\tn{T} {A}^{\gamma} \bs e_2  -\lambda
\end{bmatrix}.
\ee
To evaluate the trace and determinant of $M_{\bs \lambda}$, we first note that $\{\bs e_1,\bs e_2,\bs\lambda\}$ is an orthogonal basis for ${A}^{\gamma}$. In that basis, $A^{\gamma}$ takes the form
\be
{A}^{\gamma}=
\begin{bmatrix}
      \bs e_1^\tn{T} {A}^{\gamma} \bs e_1  & \bs e_1^\tn{T} {A}^{\gamma} \bs e_2 & 0  \\[0.3em]
      \bs e_2^\tn{T} {A}^{\gamma} \bs e_1 & \bs e_2^\tn{T} {A}^{\gamma} \bs e_2 & 0  \\[0.3em]
     \bs \lambda^\tn{T} {A}^{\gamma} \bs e_1 & \bs \lambda^\tn{T} {A}^{\gamma} \bs e_2 & \lambda
\end{bmatrix},
\ee
Let $\{\lambda_a,\lambda_b,\lambda\}$ be the three eigenvalues of $A^\gamma$. The following identities follow from the above matrix representation
\be
\tn{Tr}{A}^{\gamma}&=&\bs e_1^\tn{T} {A}^{\gamma} \bs e_1+\bs e_2^\tn{T} {A}^{\gamma} \bs e_2+\lambda=\lambda_a+\lambda_b+\lambda=0, 
\\
\det {A}^{\gamma}&=&\lambda_a\lambda_b\lambda=\lambda [( \bs e_1^\tn{T} {A}^{\gamma} \bs e_1)( \bs e_2^\tn{T} {A}^{\gamma} \bs e_2)-( \bs e_1^\tn{T} {A}^{\gamma} \bs e_2)( \bs e_2^\tn{T} {A}^{\gamma} \bs e_1)].
\ee
where we used fluid incompressibility in the first equation. From the above, we derive the following formulae
\be
\bs e_1^\tn{T} {A}^{\gamma} \bs e_1+\bs e_2^\tn{T} {A}^{\gamma} \bs e_2&=&-\lambda, \\
( \bs e_1^\tn{T} {A}^{\gamma} \bs e_1)( \bs e_2^\tn{T} {A}^{\gamma} \bs e_2)-( \bs e_1^\tn{T} {A}^{\gamma} \bs e_2)( \bs e_2^\tn{T} {A}^{\gamma} \bs e_1)&=&\lambda_a\lambda_b,
\ee
which imply the following expressions for the trace and determinant of $M$
\bse
\be
\label{eq:app_tracedetM}
\tn{Tr}M_{\bs \lambda}&=&-3\lambda, \\
\det M_{\bs \lambda} &=&2\lambda^2+\lambda_a\lambda_b.
\ee
\ese
From these expression, the eigenvalues of $M_{\bs \lambda}$ read
\be
\lambda_{\pm}^{M_{\bs \lambda}}=\f{1}{2}(\tn{Tr}M_{\bs \lambda}\pm \sqrt{\tn{Tr}M^2-4\det M_{\bs \lambda}})=
\f{1}{2}(-3\lambda\pm\sqrt{\lambda^2-4\lambda_a\lambda_b}).
\ee
Since $\lambda=-\lambda_a-\lambda_b$ this further simplifies to
\be
\label{eq:Meigs}
\lambda_{\pm}^{M_{\bs \lambda}}=
\f{1}{2}(-3\lambda\pm\sqrt{(\lambda_a-\lambda_b)^2}).
\ee

We now use the eigenvalues~(\ref{eq:Meigs}) to analyze the linear stability of the fixed points of the Jeffrey Eq.~(\ref{eq:JeffreysEq2appen}). Let's first consider the case when ${A}^{\gamma}$ has all real eigenvalues $\lambda_1<\lambda_2<\lambda_3$ with eigenvectors $\{\bs\lambda_1,\bs\lambda_2,\bs\lambda_3\}$, which are also the fixed points of Eq.~(\ref{eq:JeffreysEq2appen}). In this case, the eigenvalues of the linearized system~(\ref{eq:JeffLinearM}) are also real and Eq.~(\ref{eq:Meigs}) simplifies to
\be
\label{eq:MeigsArealeigs}
\lambda_{\pm}^{M_{\bs \lambda}}=
\f{1}{2}(-3\lambda\pm |\lambda_a-\lambda_b|).
\ee
Furthermore, since $\lambda_1+\lambda_2+\lambda_3=0$ by the incompressibility, we must have $\lambda_1=-|\lambda_1|<0$, $\lambda_3>0$ and $|\lambda_2|<\tn{min}(|\lambda_1|,\lambda_3)$. For the fixed point $\bs \lambda_1$, the eigenvalues are are always positive
\be
\lambda_{\pm}^{M_{\bs \lambda_1}}=
\f{1}{2}(-3\lambda_1\pm |\lambda_3-\lambda_2|)=\f{1}{2}[-3\lambda_1\pm (\lambda_3-\lambda_2)]
=\f{1}{2}[-3\lambda_1\pm (-\lambda_1-2\lambda_2)]>0,
\ee
implying that $\bs \lambda_1$ is a repulsive node. For the fixed point $\bs \lambda_2$, the eigenvalues are
\be
\lambda_{\pm}^{M_{\bs \lambda_2}}=
\f{1}{2}(-3\lambda_2\pm |\lambda_1-\lambda_3|)=\f{1}{2}(-3\lambda_2\pm |\lambda_2+2\lambda_3|)=\f{1}{2}[-3\lambda_2\pm (\lambda_2+2\lambda_3)].
\ee
Explicitly,
\bse
\be
\lambda_{+}^{M_{\bs \lambda_2}}=\lambda_3-\lambda_2>0, \\
\lambda_{-}^{M_{\bs \lambda_2}}=-2\lambda_2-\lambda_3=\lambda_1-\lambda_2<0.
\ee
\ese
Thus, $\bs\lambda_2$ is a saddle point. Finally, for $\bs \lambda_3$, the eigenvalues are always negative
\be
\lambda_{\pm}^{M_{\bs \lambda_3}}=
\f{1}{2}(-3\lambda_3\pm |\lambda_1-\lambda_2|)=\f{1}{2}[-3\lambda_3\pm (\lambda_3+2\lambda_2)]<0,
\ee
implying that $\bs \lambda_3$ is an attracting node.  We conclude that the asymptotically stable orientations of Eq.~(\ref{eq:JeffreysEq2appen}) for the case when ${A}^{\gamma}$ has three real eigenvalues are given by $\pm\bs \lambda_3$, since these are the only attracting fixed points on the sphere of orientations. We can estimate the characteristic time $\tau_{\bs \lambda_3}$ needed to converge onto the stable orientation $\bs \lambda_3$ as the inverse of the average of the eigenvalues $\lambda_{\pm}^{M_{\bs \lambda_3}}$
\be
\tau_{\bs \lambda_3}^{-1}=-(\lambda_{+}^{M_{\bs \lambda_3}}+\lambda_{-}^{M_{\bs \lambda_3}})/2=\f{3}{2}\lambda_3.
\ee
We now consider the case when ${A}^{\gamma}$ has a pair of complex conjugate eigenvalues and one real eigenvalue $\{\lambda_1,\lambda_1^*,\lambda_3\}$. We write the complex eigenvalue as $\lambda_1=\lambda_1^\tn{r}+i \lambda_1^\tn{i}$. In this case, the only fixed point of the Jeffrey Eq.~(\ref{eq:JeffreysEq2appen}) is given by the real eigenvector $\bs\lambda_3$. We estimate the linear stability of this fixed point. The eigenvalues of the linearized system~(\ref{eq:JeffLinearM}) become
\be
\lambda_{\pm}^{M_{\bs \lambda_3}}=
\f{1}{2}[-3\lambda_3\pm\sqrt{(\lambda_1-\lambda_1^*)^2}]=-\f{3}{2}\lambda_3\pm i|\lambda_1^\tn{i}|.
\ee
We see that, if the only real eigenvalue $\lambda_3$ is positive, then $\bs \lambda_3$ is an attractive spiral. Otherwise, it's a repulsive spiral and the asymptotic state of Eq.~(\ref{eq:JeffreysEq2appen}) is given by a stable limit cycle, to be discussed in the next section in more detail. The timescale associated with the convergence on or divergence away from $\bs \lambda_3$ is given by the absolute value of the real part of $\lambda_{\pm}^{M_{\bs \lambda_3}}$
\be
\tau_{\bs \lambda_3}^{-1}=\f{3}{2}|\lambda_3|.
\ee

\subsection{Limit cycle case}
\label{Appsec:Jeff_limitcycle}

In the case when ${A}^{\gamma}$ has complex eigenvalues $\lambda_{1,2}$ and the real eigenvalue is negative $\lambda<0$, the asymptotic solution to the Jeffrey Eq.~(\ref{eq:JeffreysEq2appen}) is given by a limit cycle. The limit cycle is the great circle perpendicular to the real eigenvector $\bs p^*$ of ${A}^{\gamma}$. To show this, we introduce the orthonormal basis
\be
\bs n_1=\bs w \times \bs p^*/\|\bs w \times \bs p^*\|, \quad \bs n_2=\bs n_1 \times \bs p^*,
\ee
where $\bs w$ is a random nonzero vector. Note that the two orthogonal vectors $\bs n_1$ and $\bs n_2$ span the plane of the great circle perpendicular to the real eigenvector $\bs p^*$. We look for solutions of the form
\be
\label{eq:limit_cycle_ansatz}
\bs p(t)=\sin \theta(t)\bs n_1+\cos \theta(t)\bs n_2.
\ee
Plugging the above ansatz into the Jeffrey Eq.~(\ref{eq:JeffreysEq2appen}) yields
\be
\label{eq:ansatz_into_Jeffrey}
\dot\theta\cos \theta\bs n_1-\dot\theta\sin \theta\bs n_2=[I-(\sin \theta\bs n_1+\cos \theta\bs n_2)(\sin \theta\bs n_1^\tn{T}+\cos \theta\bs n_2^\tn{T})](\sin \theta{A}^{\gamma}\bs n_1+\cos \theta{A}^{\gamma}\bs n_2).
\ee
To simplify the above expression, we introduce the following notation for the  submatrix of ${A}^{\gamma}$
\be
\bs M=
\begin{bmatrix}
      \bs n_1^\tn{T} A \bs n_1 & \bs n_1^\tn{T} A \bs n_2           \\[0.3em]
      \bs n_2^\tn{T} A \bs n_1 &  \bs n_2^\tn{T} A \bs n_2        
     \end{bmatrix},
\ee
and project~Eq.~(\ref{eq:ansatz_into_Jeffrey}) onto $\bs n_1$ and $\bs n_2$
\be
\dot\theta\cos \theta(t)&=&
\sin \theta M_{11}+\cos \theta M_{12}-\sin\theta(\sin^2\theta M_{11}+\sin\theta\cos\theta M_{12}+\cos\theta\sin\theta M_{21}+\cos^2\theta M_{22}), \\
\dot\theta\sin \theta(t)&=&-\sin\theta M_{21}-\cos\theta M_{22}+
\cos\theta(\sin^2\theta M_{11}+\sin\theta\cos\theta M_{12}+\cos\theta\sin\theta M_{21}+\cos^2\theta M_{22}).
\ee
We combine the two equations into a single one by taking a linear combination with weights $\cos\theta$ and $\sin\theta$
\be
\dot\theta\cos^2 \theta(t)+\dot\theta\sin^2 \theta(t)=\dot\theta=\cos\theta\sin \theta M_{11}+\cos^2 \theta M_{12}-\sin^2\theta M_{21}-\sin\theta\cos\theta M_{22},
\ee
which further simplifies to
\be
\dot\theta=
\f{M_{11}-M_{22}}{2}\sin 2\theta +\f{M_{12}+M_{21}}{2}\cos 2\theta +\f{M_{12}-M_{21}}{2}.
\ee
Introducing $A=M_{11}-M_{22}$, $B=M_{12}+M_{21}$ and $C=M_{12}-M_{21}$, we obtain
\be
\label{eq:theta_dot}
2\dot\theta=
A\sin 2\theta +B\cos 2\theta +C.
\ee
This is a first-order nonlinear differential equation. Since the nonlinear term is smooth, the unique (up to the $2\pi$ period) solution exists, which validates the ansatz~(\ref{eq:limit_cycle_ansatz}) and proves the existence of a limit cycle.

We now explicitly calculate the period $T$ of the limit cycle. To this end, integrate Eq.~(\ref{eq:theta_dot}) over $T$
\be
4\pi&=&CT+\int_0^T (A\sin 2\theta +B\cos 2\theta)dt \\
&=&
CT+\int_0^{2\pi} (A\sin 2\theta +B\cos 2\theta)\f{1}{\dot\theta}d\theta \\
&=&
CT+2\int_0^{2\pi}(1-\f{C}{A\sin 2\theta +B\cos 2\theta +C})d\theta.
\ee
We get the following equation for $T$
\be
T=\int_0^{2\pi}\f{2}{A\sin 2\theta +B\cos 2\theta +C}d\theta.
\ee
This expression can be expressed as
\be
T=
2\int_0^{2\pi}\f{1}{A\sin \theta' +B\cos \theta' +C}d\theta'=
2\int_0^{2\pi}\f{1}{\sqrt{A^2+B^2}\sin \hat{\theta} +C}d\hat{\theta},
\ee
where we changed variables twice using $\theta'=2\theta$ and $\hat{\theta}=\theta'+\ga$, where $\sin\ga=B/\sqrt{A^2+B^2}$ and we used the periodicity of the integrand to keep the integration limit as $[0,2\pi)$. The final integral can be evaluated using contour integration. We first change the variables $z=\sqrt{A^2+B^2}e^{i\theta}$, which yields
\be
T=4\oint \f{dz}{z^2+2iCz-(A^2+B^2)}.
\ee
We note that $C^2>A^2+B^2$ corresponds to $A$ having complex eigenvalues, which is the case of interest. In this case, the integrand has one simple pole inside the integration contour (circle of radius $\sqrt{A^2+B^2}$ centered at the origin) given by one of the roots of the integrand denominator. Applying the residue theorem, yields
\be
T=\f{4\pi}{\sqrt{C^2-A^2-B^2}}.
\ee
This can be related to the original matrix $A$ and its negative real eigenvalue $\lambda$ as
\be
T=\f{4\pi}{\sqrt{2 \det A/\lambda+\lambda^2-\tn{tr}(A^2)}}.
\ee
Assuming the complex eigenvalues take the form $\lambda_{1,2}=\alpha \pm i \beta$, this further simplifies to
\be
T=\f{2\pi}{\beta}.
\ee
Therefore, the angular frequency of the limit cycle is given by the imaginary part of the complex eigenvalue. These results agree with the analysis in~\cite{Bretherton1962} obtained using a different method.

\subsection{Velocity gradient of the Stokes flow around a sphere}
\label{Appsec:AStokesFlow}
In this section, we compute the velocity gradient in Eq.~(\ref{eq:VGspherical_ortho}) due to the Stokes flow around a sinking particle. We first carry out the calculation in the curvilinear orthogonal coordinate basis $\{\p_r,\p_\theta,\p_\phi\}$ with the metric tensor $g_{ij}=\tn{diag}(1,r^2,r^2\sin^2\theta)$ and then transform to the usual orthonormal system $\{\hat{\bs r},\hat{\bs\theta},\hat{\bs\phi}\}$. The transformation between the two systems is encoded in the Jacobian
\be
\label{eq:Jacobian}
J=\tn{diag}(1,r,r\sin\theta).
\ee

In the curvilinear system $\{\p_r,\p_\theta,\p_\phi\}$, the Stokes flow~(\ref{eq:Stokes_flow_sphere}) reads
\be
v=v^r\p_r+v^\theta \p_\theta=U\cos\theta\Big(1+\f{R^3}{2r^3}-\f{3R}{2r}\Big)\p_r+
U\sin\theta\Big(-\f{1}{r}+\f{R^3}{4r^4}+\f{3R}{4r^2}\Big)\p_\theta.
\ee
To compute the velocity gradient (1,1)-tensor $A_{ij}=\nabla_j v^i$, we note that the only nonzero  Christoffel symbols are
\be
&&
\Gamma^{r}_{\theta\theta}=-r,\quad 
\Gamma^{r}_{\phi\phi}=-r\sin^2\theta, \\
&&
\Gamma^\theta_{r\theta}=\Gamma^\theta_{\theta r}=1/r, \quad
\Gamma^\theta_{\phi\phi}=-\sin\theta\cos\theta, \\
&&
\Gamma^\phi_{r\phi}=\Gamma^\phi_{\phi r}=1/r, \quad \Gamma^\phi_{\theta\phi}=\Gamma^\phi_{\phi \theta}=\cot\theta.
\ee
Using covariant differentiation, we find the velocity gradient tensor components (in the $\{\p_r,\p_\theta,\p_\phi\}$ basis)
\bse
\be
\nabla_r v^r&=&\p_r v^r, \quad \nabla_r v^\theta=\p_r v^\theta+v^\theta/r, \quad \nabla_r v^\phi=0,\\
\nabla_\theta v^r&=&\p_\theta v^r-r v^\theta, \quad \nabla_\theta v^\theta=\p_\theta v^\theta+v^r/r, \quad \nabla_\theta v^\phi=0,\\
\nabla_\phi v^r&=&0,\quad \nabla_\phi v^\theta=0, \quad\nabla_\phi v^\phi=v^r/r+\cot\theta v^\theta.
\ee
\ese
Explicit calculation gives the following expressions for the tensor entries
\bse
\be
&&\nabla_r v^r=U\cos\theta\Big(-\f{3R^3}{2r^4}+\f{3R}{2r^2}\Big), \\
&&\nabla_r v^\theta=U\sin\theta\Big(\f{1}{r^2}-\f{R^3}{r^5}-\f{3R}{2r^3}\Big)+U\sin\theta\Big(-\f{1}{r^2}+\f{R^3}{4r^5}+\f{3R}{4r^3}\Big)
=
U\sin\theta\Big(-\f{3R^3}{4r^5}-\f{3R}{4r^3}\Big)
, \\
&&\nabla_r v^\phi=0, \\
&&\nabla_\theta v^r=-U\sin\theta\Big(1+\f{R^3}{2r^3}-\f{3R}{2r}\Big)+U\sin\theta\Big(1-\f{R^3}{4r^3}-\f{3R}{4r}\Big)=
U\sin\theta\Big(-\f{3R^3}{4r^3}+\f{3R}{4r}\Big)
, \\
&&\nabla_\theta v^\theta=U\cos\theta\Big(-\f{1}{r}+\f{R^3}{4r^4}+\f{3R}{4r^2}\Big)+U\cos\theta\Big(\f{1}{r}+\f{R^3}{2r^4}-\f{3R}{2r^2}\Big)=
U\cos\theta\Big(\f{3R^3}{4r^4}-\f{3R}{4r^2}\Big)
, \\
&&\nabla_\theta v^\phi=0, \\
&&\nabla_\phi v^r=0,\\
&&\nabla_\phi v^\theta=0,\\
&&\nabla_\phi v^\phi=U\cos\theta\Big(\f{1}{r}+\f{R^3}{2r^4}-\f{3R}{2r^2}\Big)+U\cos\theta\Big(-\f{1}{r}+\f{R^3}{4r^4}+\f{3R}{4r^2}\Big)=
U\cos\theta\Big(\f{3R^3}{4r^4}-\f{3R}{4r^2}\Big)
.
\ee
\ese
As a sanity check, we compute the flow divergence
\be
\nabla_r v^r+\nabla_\theta v^\theta+\nabla_\phi v^\phi=0,
\ee
which vanishes, as expected. In the matrix form, the above tensor reads ($U=1$ and $R=1$)
\be
A_{ij}=\nabla_j v^i=
\begin{bmatrix}
      \big(-\f{3}{2r^4}+\f{3}{2r^2}\big) \cos\theta & \big(-\f{3}{4r^3}+\f{3}{4r}\big)\sin\theta & 0           \\[0.3em]
       \big(-\f{3}{4r^5}-\f{3}{4r^3}\big)\sin\theta & \big(\f{3}{4r^4}-\f{3}{4r^2}\big)\cos\theta           & 0 \\[0.3em]
       0           & 0 & \big(\f{3}{4r^4}-\f{3}{4r^2}\big)\cos\theta
     \end{bmatrix}.
\ee
Finally, we use the Jacobian J~[Eq.~(\ref{eq:Jacobian})] to express $A$ in the orthonormal basis $\{\hat{\bs r},\hat{\bs\theta},\hat{\bs\phi}\}$
\be
J A J^{-1}=
\f{3}{4} (\f{1}{r^2}-\f{1}{r^4}) \cos\theta 
\begin{bmatrix}
     2 & \tan\theta & 0            \\[0.3em]
       - \f{r^2+1}{r^2-1} \tan\theta & -1 & 0  \\[0.3em]
      0 & 0 & -1
     \end{bmatrix},
\ee
which yields Eq.~(\ref{eq:VGspherical_ortho}), in agreement with the calculation in~\cite{Kiorboe1999} where the velocity gradient was computed using a different method.

\subsection{Ellipsoids in the Stokes flow: stagnation lines and particle surface}
\label{sec:Stokesflow_asymp_critical}

The eigenvalues of the velocity gradient $A$~[Eq.~(\ref{eq:VGspherical_ortho})] on the stagnation line ($\theta=0,\pi$) and the particle surface ($r=1$) have multiplicity greater than one. In this case, the analysis of Section~\ref{sec:Jeffrey_asymp} does not directly apply, yet these special locations will be important for the encounter process of non-motile microorganisms, which can only approach the sinking particle near the stagnation line $\theta=\pi$. On the stagnation lines $\theta=0,\pi$, Eq.~(\ref{eq:VGspherical_ortho}) reduces to
\be
\label{eq:VGspherical_ortho_stagnation}
A_{ij}(r,\theta=0,\pi,\phi)=
\pm\f{3}{4}\Big(\f{1}{r^2}-\f{1}{r^4}\Big)
\begin{bmatrix}
       2 & 0 & 0           \\
       0 &  -1           & 0 \\
       0           & 0 &  -1
\end{bmatrix},
\ee
where $+/-$ corresponds to $\theta=0$ and $\theta=\pi$, respectively. This simple diagonal structure implies that on the upstream stagnation line ($\theta=\pi$) rods align tangentially to the sinking particle, while on the downstream stagnation line ($\theta=0$), rods align vertically. This picture can be inferred from Fig.~\ref{fig:particles_fixed_StokesFlow}(b) by taking the limit $\rho\to 0$. Since $A^{\gamma=-1}=-A^\tn{T}$, we immediately obtain the response of disks. Disks align tangentially to the particle surface for $\theta=\pi$ (with axis of symmetry in the vertical direction), while they lie in the $r-\theta$ plane for $\theta=0$. Therefore, non-motile rods or disks approaching the sinking particle along the $\theta=\pi$ stagnation line orient with their longer dimension tangential to the particle surface. We now compute $A$ on the particle surface to see if shear tends to maintain such a tangential orientation. At $r=1$, the only nonzero component of the velocity gradient is $A_{\theta r}(r=1,\theta,\phi)=-\f{3}{2}\sin\theta$. This structure implies that the tangential orientations of rods (symmetry axis along $\theta-\phi$) and disks (symmetry axis along $r$) are the null vectors. Thus, to zeroth order, shear maintains the tangential orientation of rods and disks as they are advected around the sinking particle. 

\subsection{Methods: numerical simulations}
\label{appsec:sim_methods}

\textbf{Time stepping.} To numerically integrate the ballistic model~[Eq.~(\ref{eq:EOMbacterium})], we discretized the equations of motion using the classical Runge--Kutta method (RK4). Depending on the sinking speed, the time-step was chosen between $\Delta t=0.075 \tau_\tn{b}$ for $U\sim U_\tn{b}$ and $\Delta t=0.005 \tau_\tn{b}$ for $U\sim 100 U_\tn{b}$, where $\tau_\tn{b}=l_\tn{b}/U_\tn{b}$ is the time needed for the bacterium to travel distance equal to its bodylength. To integrate the quasi-ballistic model with rotational diffusion~[Eq.~(\ref{eq:EOMdiff})], we used the stochastic version of the Euler method; at each time step, the diffusive term is discretized by sampling a $3\times 1$ vector with normally distributed entries with zero mean and variance $2 D_\tn{r}\Delta t$. With the rotational diffusion coefficient $D_\tn{r}=0.25 \tn{s}^{-1}$, the time step varied between $0.04-0.2\tn{ms}$.

\textbf{Estimation of the encounter efficiency.} For a given sinking particle size $R$ and sinking speed $U$, we estimated the encounter efficiency by discretizing Eq.~(\ref{eq:eta}). We typically sampled the encounter probability $P(\rho)$ on a non-uniform grid to resolve the accumulation of $P(\rho)$ near the accessibility region for slowly sinking particles or near the centerline for fast sinking particles~(see Figs.~\ref{fig:encounter_prob_heatmap} and~\ref{fig:EC_detailed}); the number of points on the $\rho$-grid was always at least 50 for motile bacteria and 10 for non-motile bacteria. Once the estimate of $P(\rho)$ had been obtained, the integral in Eq.~(\ref{eq:eta}) was evaluated using the trapezoidal rule.

To estimate the encounter probability $P(\rho)$ starting in the initial plane $z=-6R$ at distance $\rho$ away from the centerline for random initial orientations, we considered an ensemble of initial orientations by sampling along a spherical spiral. Such a sampling gives an approximately uniform distribution of points on a unit sphere of orientations. By using spherical spiral to sample initial orientations rather than choosing them randomly~(that is, choosing $3\times 1$ vectors with normal entries with zero mean and normalizing them to unit length), we obtained faster convergence by avoiding random clustering of points on the unit sphere. The number of initial orientations was chosen high enough to guarantee that the solid angle the sinking particle extended at the initial bacterial location contained at least five initial orientations~(the number of blue dots in the inset in Fig.~\ref{fig:sample_trajectories} was always at least five). For such an angular resolution, the number of initial orientations varied between $O(10^2)$ for fast sinking particles up to $O(10^4)$ for slowly sinking particles, for which the accessibility region with $P(\rho)>0$ was largest. In general, to estimate the encounter efficiency $\eta$ for a given $(R,U)$ pair, we simulated about $O(10^4)$ trajectories for fast sinking particles and up to $O(10^6)$ trajectories for slowly sinking particles. In total, due to scanning the $(R,U)$ parameter space in different shear ON/OFF configurations, the this work summarizes the results of simulating about $O(10^8)$ bacterial trajectories.

\textbf{Interception criterion.} The sinking particle was assumed to be a perfect absorber: geometric overlap between any part of a bacterium and the particle was counted as an encounter. In simulations, for simplicity, we computed this geometric overlap by approximating elongated bacteria~($\ga>1$) of length $l_\tn{b}$ by a cylinder with spherical caps. The cylinder length, including caps, is $l_\tn{b}$ and its width is $l_\tn{b}/\ga$. With this simplification, determining the interception is equivalent to determining the distance between the cylinder centerline and the particle center. Similarly, the geometry of oblate particles~($\ga<1$) was approximated by considering four cylinders~(with spherical caps) of length $l_\tn{b}$ and width $l_\tn{b}\ga$. The centerlines of the cylinders lie in a plane, the centerline midpoints coincide and the centerlines are rotated at angle $\ang{45}$ - the four cylinders form two crosses rotated by $\ang{45}$. With this simplification, determining the interception is equivalent to determining the distance between the four cylinder centerlines and the particle center.

\textbf{Estimation of the distribution of interception locations.} To estimate $\xi(\theta)$ for a given $(R,U)$ pair, we considered the ensemble of the endpoints of trajectories that resulted in the interception. As described above, this ensemble resulted from scanning the $\rho$-range, the initial position at distance $\rho$ away from the particle centerline in the initial $z=-6R$-plane, as well as uniform initial orientations. This ensemble yielded a histogram of the interception colatitudes $\theta$. During construction of this histogram, the counts for each scanned position $\rho$ were further weighted by $\rho$ and the $\rho$-grid spacing, to account for the number of initial positions at distance $\rho$ being proportional to $\rho$~(circles of radius $\rho$) as well as the non-uniformity of the $\rho$-grid. Such prepared histogram of $\theta$-counts~(30 bins, bin width $\ang{6}$), normalized to a probability density function over a unit sphere, was used as the estimate of $\xi(\theta)$.

\subsection{Methods: experiments}
\label{appsec:exp_methods}

\textbf{Cell cultured.} \textit{Phaeodactylum tricornutum} cells (strain CCMP2561) were cultured in f/2 medium (Guillard and Ryther 1962) mixed with artificial seawater. Artificial seawater was prepared by dissolving $\SI{35}{\gram}$ of artificial sea salt (Instant Ocean, Spectrum Brands) in $\SI{1}{\liter}$ double distilled water, filtered through a $\SI{0.2}{\micro\meter}$ filter and autoclaved. Cultures were propagated in $\ang{18}$ C in AlgaeTron AG 230 PSI (Photon Systems Instruments) with 14h/10h light/dark cycle. For the experiments, cells in the exponential growing phase were used. Cell length and width were $21.2\pm2.4~\SI{}{\micro\meter}$ ($n=14$) and $3.12 \pm 0.57~\SI{}{\micro\meter}$ ($n=14$), respectively, as measured by phase microscopy.

\textbf{Experimental procedure.} Alginate beads were prepared using a mix of sodium alginate salt from brown algae (1.5\% w/v, medium viscosity; Sigma) with 50~mM ethylenediamine tetra acetic acid (EDTA) in double distilled water (DDW). Beads were prepared by dripping the alginate solution from a 1~mL syringe at rate of $\SI{60}{\micro\liter\per\minute}$ from a height of 20~cm to beaker containing 0.5~M $\tn{CaCl}_2$ in DDW. The $\tn{CaCl}_2$ solution was stirred at 300~rpm, using a magnetic stir-bar. Flow dynamics were studied in microfluidic chip (Sticky-Slide 0.4 – IBIDI). For the experiment, single bead was trapped at the center of the channel using a glass cover slide. A syringe pump (Harvard PHD2000) was then used to feed the channels with artificial sea water at the desired flow speed ($\SI{160}{\micro\meter\per\second}$). Chanel was visualized using a Nikon (eclipse TI-2) microscope at magnification of $4\times 1.5\times 10$ ($60\times$) and 20~fps using Orca flash 4.0 (Hamamatsu) camera.   

\textbf{Image analysis.} Image analysis was performed on 50 consecutive images using ImageJ (Rueden, C. T et al. (2017)). In general $1098 \pm 50$ cells/image were measured. To extract the orientation from each cell, median image intensity was calculated using \lq Stacks/Z Projection\rq~function and subtract from all images. \lq FFT/Bang pass\rq~filter was used with small cutoff of 1~pxl and big cutoff of 10~pxl. \lq Minimum\rq~filter was used with 1~pxl cutoff. Data was transformed to binary using \lq Make binary\rq~function with default parameters. Finally, \lq Analyze particle\rq~function (Particle 30-1000~pxl) was used to collect the orientation data.

\bibliographystyle{unsrt}
\bibliography{thebibliography}

\end{document}